\documentclass[showpacs,amsmath,amssymb,prd,nofootinbib]{revtex4}
 
\usepackage{graphicx}% Include figure files
\usepackage{dcolumn}% Align table columns on decimal point
\usepackage{bm}% bold math
\usepackage{epsf}
\usepackage{hyperref}
\usepackage{amsfonts}
\usepackage{mathrsfs}

\begin{document}
 
\title{Mode coupling mechanism for late--time Kerr tails}

\author{Lior M.~Burko$^{1,2}$ and Gaurav Khanna$^{3}$}

\affiliation{
$^1$ Department of Physics, Chemistry, and Mathematics, Alabama A\&M University, Normal, Alabama 35762\\ 
$^2$ Theiss Research, La Jolla, California 92037\\ 
$^3$ Department of Physics and Center for Scientific Computing and Visualization Research, University of Massachusetts Dartmouth, North Dartmouth, Massachusetts 02747 } 

\date{December 18, 2013}

\begin{abstract}
We consider the decay rate for scalar fields in Kerr spacetime. We consider pure initial (azimuthal) multipoles $\ell'$ with respect to the class which includes Boyer--Lindquist coordinates, and focus attention on the decay rate of the multipole $\ell$. We use an iterative method proposed by Gleiser, Price, and Pullin, and identify the mode coupling mechanism through the iterations in powers of the square of the Kerr black hole's specific angular momentum that gives rise to a decay rate formula recently proposed by Zengino\u{g}lu, Khanna, and Burko. Modes $\ell$ may be excited through different channels, each leading to its own decay rate. The asymptotic decay rate of the mode $\ell$ is the slowest of the decay rate of the various channels. In some cases, more than one channel leads to the same decay rate, and then the amplitude of the mode is the sum of the amplitudes of the partial fields generated by the individual channels. We also show that one may identify the asymptotically--dominant channel of mode excitations, and obtain approximate results for the mode of interest by studying the dominant channel. The results of the dominant channel approximation approach the full--mode results at late times, and their difference approaches zero quadratically in inverse time.
\end{abstract}

 
\maketitle

\section{Introduction and summary}

The decay rate of late--time tails in black hole spacetime has been at the focus of much interest. The late--time decay rate in Schwarzschild spacetime is well understood \cite{price,leaver,barack1,dafermos}, and was confirmed numerically both in linearized \cite{gundlach} and fully nonlinear \cite{gundlach2,burko-ori} setups. 
Of current interest is the detailed behavior of perturbation  fields of spinning, or Kerr black holes for scalar fields \cite{krivan,barack-ori,hod1,poisson,BK-03,scheel,tiglio,ZT,BK-09,BK-11,racz,jasiulek} and higher spin fields \cite{barack,hod2,harms}. 
Very recently,  a modification to the formula that described the late--time decay rate for scalar fields was proposed  by Zengino\u{g}lu, Khanna, and Burko (ZKB) \cite{zenginoglu-khanna-burko}. Denoting by $\ell'$ the multipole moment of the initial perturbation and by $\ell$ the multipole moment of the mode of interest, it is known that the late--time behavior of the field is given by $\psi\sim t^{n}$. 
Based on (2+1)D numerical simulations, ZKB proposed that with the exception of the case in which $\ell'$ is the slowest decaying mode (for which case the decay rates are given by $-n=\ell'+\ell+3$), all other azimuthal modes Ñeven or oddÑ appear to decay along $r={\rm const}$ according to $-n=\ell'+\ell+1$.\footnote{We are interested in this Paper mostly in the tail behavior along $r={\rm const}$, approaching future timelike infinity $i^+$. ZKB also found the decay rate along future null infinity $\mathscr{I^+}$, specifically, $-n^{\mathscr{I^+}}=\ell+2$ if $\ell\ge\ell'$, and $-n^{\mathscr{I^+}}=\ell$ if $\ell\le\ell'-2$.}

The ZKB proposal differs from other  numerically based predictions in several key elements. Specifically, focusing on the azimuthal modes ($m=0$), in \cite{BK-11} it was proposed that 
$-n=\ell'+\ell+3$ for $\ell\ge \ell'$ and $-n=\ell'+\ell+1$ for $\ell<\ell'$. The ZKB formula coincides with the formula in \cite{BK-11}  for all cases that $\ell<\ell'$ (``down excitations"). However, for cases with $\ell\ge \ell'$ (``up excitations") the two formulas agree only when $\ell'$ is the lowest excitable mode, that is if $\ell'=0,1$. 
Specifically, ZKB find a slower decay rate for ``up" excitations. The main difference in the numerical work of ZKB and \cite{BK-11}  is the technological code development, specifically the double hyperboloidal layers \cite{hyper} and compactification at both infinity and at the horizon, that allowed ZKB to run to much later times than in \cite{BK-11}. It is the much later evolutions that allowed ZKB to identify certain decay rates as intermediate rather than asymptotic.  More recently, Spilhaus and Khanna further developed the (2+1)D numerical technology, and confirmed the numerical results of ZKB for more values of the multipoles, up to $\ell',\ell=16$ \cite{khanna-spilhaus}. 

ZKB further proposed, without proof, a mechanism for mode coupling that produces the numerically observed decay rates. However, the (2+1)D approach, that was so effective in determining the asymptotic decay rate, is not very effective in determining the mode coupling mechanism. In this Paper we adopt a different approach, that allows us to (i) independently check the ZKB decay--rate formula, and (ii) identify and corroborate the mode coupling mechanism that explains the ZKB formula. 

The approach we undertake is the approach first proposed by Gleiser, Price, and Pullin (hereafter, GPP) \cite{GPP}. GPP expand the field and the wave operator in powers of $(a/M)^2$, where $M$ is the Kerr black hole's mass and $a$ is its specific spin angular momentum, and solve the field equations iteratively. The greatest benefit of the GPP approach is that the principal symbol of the resulting partial differential equations is just that of wave evolution in Schwarzschild spacetime, so that one may readily cast the field equations in (1+1)D, using, {\em e.g.}, the characteristic methods that proved to be very effective for Schwarzschild wave evolution. The iterations couple modes of different multipoles $\ell$, and the effective potential at each iteration depends only on the multipole of the mode for which one solves. The coupling to other modes takes place only in the source term of the individual mode equations. These great benefits of the GPP approach are met with a practical difficulty: for each initial mode -- final mode pair of interest one needs to write the field equations of the various modes {\em ad hoc}. Therefore, the elegance of the approach and the simpler numerical framework are offset by the analytical derivation of the equations to solve. Notably, the evolution of any multipole mode $\ell$ can be found accurately using modern (2+1)D approaches, such as those used recently by \cite{racz,harms,zenginoglu-khanna-burko,khanna-spilhaus}. The benefit of the GPP approach that we employ here is that it allows us to track the intricate mode coupling mechanism in order to better understand the results proposed by ZKB, and determine the decay rate of sub-dominant modes that would  become dominant at asymptotically late times. 

The source term in each GPP partial differential equation is a sum of terms that typically involve several modes of lower--order in the iterative scheme. One may view each inhomogeneous equation as an effective equation that sums the contributions of excitations through various channels, each channel going through a different route in an iteration order -- multipole chart. There are multiple channels through which a particular mode is excited. Each channel leads to a particular asymptotic decay rate. 
Sometimes more than just one channel leads to the same asymptotically--dominant decay rate, in which case all such channels need to be considered simultaneously. In other cases there is a unique asymptotically--dominant channel. Understanding the general features of the mode coupling mechanism allows us also to identify the dominant channel of excitation. We propose an approximation to the full mode--coupling scheme that utilizes the dominant channel only, neglecting asymptotically sub-dominant channels. We justify the use of the dominant channel approximation and show that its results approach asymptotically the results of the full--mode scheme. The latter approach makes the GPP formalism from the computational point of view a very efficient one. 

This Paper is organized as follows. In Section \ref{sec:gpp} we review the GPP formalism, in Section \ref{full} we apply the GPP formalism to a specific explicit case, specifically the case $\ell'=2$, and in Section \ref{dominant} we propose the dominant channel approach, and apply it to the cases $\ell'=2,4$, and $6$. 

\section{The Gleiser--Price--Pullin (GPP) Formalism}\label{sec:gpp}

The Kerr metric in Boyer--Lindquist coordinates $(t,r,\theta,\varphi)$
reads
\begin{eqnarray} \label{bl_metric} 
\,ds^2_{\rm{BL}} &=& -\left(1-\frac{2Mr}{\Sigma}\right) dt^2 - \frac{4 a
  M r}{\Sigma}\sin^2\theta \,dt\,d\varphi + \frac{\Sigma}{\triangle}\,dr^2 \nonumber \\ 
&+& \Sigma \,d\theta^2+\left(r^2+a^2+\frac{2Ma^2r\sin^2\theta}{\Sigma}\right)\,\sin^2\theta\,d\varphi^2,\end{eqnarray}
where $\Sigma := r^2 + a^2 \cos\theta^2$, and $\triangle :=r^2 + a^2 -
2Mr$. We denote the mass of the Kerr spacetime by $M$, and its specific angular momentum by $a$. We use geometrized units in which $G=c=1$. 
%It is common to introduce the tortoise coordinate by $$ \label{tortoise} \frac{dr_*}{dr} = \frac{r^2+a^2}{\Delta} .$$

The original Boyer--Lindquist radial coordinate $r$ is uncomfortable to work with numerically in the GPP framework, as certain source terms diverge approaching the EH, rendering corresponding evolution equations divergent. 
Following GPP we introduce a new radial coordinate $\rho(r,a)$, defined by 
\begin{equation}\label{rho_def}
r:=M+\sqrt{\rho^2-2M\rho+M^2-a^2}\, ,
\end{equation}
for which the event horizon (EH) is always located at $\rho_{\rm EH}=2M$, independently of $a$.  In Appendix \ref{app_a} we list some of the mathematical properties of working with the $\rho$ coordinate. 

We next specialize to even modes (odd modes are not excited by an even initial data mode), although a similar framework can readily be written for odd modes too. The scalar--field ($s=0$) azimuthal ($m=0$) wave equation is given by the Teukolsky equation ${\cal L}(t,\rho,\theta)[\psi]=0$, where the Teukolsky operator 
\begin{eqnarray}
{\cal L}(t,\rho,\theta) &=& 
\left[\frac{\left(\rho^2-2M\rho+2M^2+2M\sqrt{\rho^2-2M\rho+M^2-a^2}\right)^2}{\rho^2\left(1-\frac{2M}{\rho}\right)}-a^2\,\sin^2\theta\right]\,\frac{\,\partial^2}{\,\partial t^2}\nonumber
\\
&-&\frac{\left(1-\frac{2M}{\rho}\right)}{\left(1-\frac{M}{\rho}\right)^2}\left(\rho^2-2M\rho+M^2-a^2\right)\,\frac{\,\partial^2}{\,\partial \rho^2}\nonumber\\
&-&\frac{1}{\rho}\,\left[\frac{(1-\frac{2M}{\rho})}{\left(1-\frac{M}{\rho}\right)^3}\,a^2+2\,\frac{\rho^2-2M\rho+M^2-a^2}{1-\frac{M}{\rho}}\right]\, \frac{\,\partial}{\,\partial \rho}
-\frac{1}{\,\sin\theta}\frac{\,\partial}{\,\partial \theta}
\left( \,\sin\theta \frac{\,\partial}{\,\partial \theta}
\right)\, .
\end{eqnarray}

Following GPP we expand the field in powers of $a^2$ to order $N$ as
\begin{equation}
\psi=\sum_{n=0}^{N}\left(\frac{a}{M}\right)^{2n}\,\psi^{(2n)}+\cdots \, ,
%\psi=\psi^{(0)}+\left(\frac{a}{M}\right)^2\psi^{(2)}+\left(\frac{a}{M}\right)^4\psi^{(4)}+\left(\frac{a}{M}\right)^6\psi^{(6)}+\left(\frac{a}{M}\right)^8\psi^{(8)}+\left(\frac{a}{M}\right)^{10}\psi^{(10)}+\left(\frac{a}{M}\right)^{12}\psi^{(12)}+\cdots
\end{equation}
and the Teukolsky operator is correspondingly expanded as
\begin{equation}
{\cal L}=\sum_{n=0}^{N}\left(\frac{a}{M}\right)^{2n}\,L^{(2n)}+\cdots \, .
%L=L^{(0)}+\left(\frac{a}{M}\right)^2L^{(2)}+\left(\frac{a}{M}\right)^4L^{(4)}+\left(\frac{a}{M}\right)^6L^{(6)}+\left(\frac{a}{M}\right)^8L^{(8)}+\left(\frac{a}{M}\right)^{10}L^{(10)}+\left(\frac{a}{M}\right)^{12}L^{(12)}+\cdots \, .
\end{equation}
The iterations satisfy 
%(for even modes; odd modes can be done analogously)
\begin{equation}\label{iterative_equation}
\sum_{n=0}^{N}L^{(2n)}\left[\psi^{(2N-2n)}\right]=0\, ,
\end{equation} 
where $N$ is the iteration order (such that we have $N+1$ evolution equations), 
or explicitly to $N=6$
\begin{eqnarray}
L^{(0)}[\psi^{(0)}]&=&0\label{L0}\\
L^{(0)}[\psi^{(2)}]&=&-L^{(2)}[\psi^{(0)}]\label{L2}\\
L^{(0)}[\psi^{(4)}]&=&-L^{(2)}[\psi^{(2)}]-L^{(4)}[\psi^{(0)}]\label{L4}\\
L^{(0)}[\psi^{(6)}]&=&-L^{(2)}[\psi^{(4)}]-L^{(4)}[\psi^{(2)}]-L^{(6)}[\psi^{(0)}]\label{L6}\\
L^{(0)}[\psi^{(8)}]&=&-L^{(2)}[\psi^{(6)}]-L^{(4)}[\psi^{(4)}]-L^{(6)}[\psi^{(2)}]-L^{(8)}[\psi^{(0)}]\label{L8}\\
L^{(0)}[\psi^{(10)}]&=&-L^{(2)}[\psi^{(8)}]-L^{(4)}[\psi^{(6)}]-L^{(6)}[\psi^{(4)}]-L^{(8)}[\psi^{(2)}]\nonumber \\
&-&L^{(10)}[\psi^{(0)}]\label{L10}\\
L^{(0)}[\psi^{(12)}]&=&-L^{(2)}[\psi^{(10)}]-L^{(4)}[\psi^{(8)}]-L^{(6)}[\psi^{(6)}]-L^{(8)}[\psi^{(4)}]\nonumber \\
&-&L^{(10)}[\psi^{(2)}]-L^{(12)}[\psi^{(0)}]\label{L12}\, ,
\end{eqnarray}
and the differential operators $L^{(2n)}$ are listed in Appendix \ref{app_operators}. 
The main properties of these operators are that, as pointed out by GPP, only the $L^{(2)}$ operator depends on the angular coordinate $\theta$ (and therefore is the only operator that is responsible for the coupling of different multipoles), that only $L^{(0)}$ includes the angular differential operator, and that all operators $L^{(2n)}$ with $n\ge 2$ are homogeneous temporal differential operators.

We write each mode $\psi^{(2n)}_{\ell}$ as $\psi^{(2n)}_{\ell}=\frac{1}{\rho}\;f^{(2n)}_{\ell}(t,\rho)\;P_{\ell}(\cos\theta)$. The operator $L^{(0)}[\psi^{(2n)}_{\ell}]$ is then given by
\begin{equation}
L^{(0)}\left[\frac{1}{\rho}f^{(2n)}_{\ell}P_{\ell}(\cos\theta)
\right] = \left\{
\frac{\rho}{1-\frac{2M}{\rho}}\,\Box f^{(2n)}_{\ell}+\frac{1}{\rho}\left[\ell (\ell+1)+\frac{2M}{\rho}\right]f^{(2n)}_{\ell}\right\}P_{\ell}(\cos\theta)\, ,
\end{equation}
where we denote $\Box:=\frac{\,\partial^2}{\,\partial t^{{2}}}-\frac{\,\partial^2}{\,\partial \rho_{*}^{2}}$ (see Appendix \ref{app_a}).  We further write
\begin{equation}
L^{(2)}\left[\frac{1}{\rho}\,f^{(2n)}_{\ell}\right]\equiv -\frac{M^2}{\rho}\,\sin^2\theta\,\frac{\,\partial^2 f^{(2n)}_{\ell}}{\,\partial t^2}+\frac{M^2}{\rho}\frac{1}{1-\frac{2M}{\rho}}\,\Delta^2\,f^{(2n)}_{\ell}\, ,
\end{equation}
where the second--order, inhomogeneous differential operator $\,\Delta^2$ is given by
\begin{eqnarray}
\Delta^2 &:=& \sum_{j=1}^{4}\Delta^2_{(j)} =
 - \frac{2M}{\rho}\frac{1}{1-\frac{M}{\rho}}\,\partial^2_t
+\frac{1}{\left(1-\frac{M}{\rho}\right)^2}\,\partial^2_{\rho_*}
-\frac{1}{\rho}\frac{1-\frac{2M}{\rho}}{\left(1-\frac{M}{\rho}\right)^3}\,\partial_{\rho_*}
\nonumber\\
&+&\frac{1}{\rho^2}\left(1-\frac{4M}{\rho}+\frac{2M^2}{\rho^2}\right)\,\frac{1-\frac{2M}{\rho}}{\left(1-\frac{M}{\rho}\right)^3} \, ,
\label{Delta2_def}
\end{eqnarray}
and define 
\begin{eqnarray}
D_4:=-\frac{1}{2}\left(\frac{M}{\rho}\right)^5\frac{1-\frac{2M}{\rho}+\frac{2M^2}{\rho^2}}{\left(1-\frac{M}{\rho}\right)^3}\, ,
\end{eqnarray}

\begin{eqnarray}
D_6:=-\frac{1}{4}\left(\frac{M}{\rho}\right)^7\frac{1-\frac{2M}{\rho}+\frac{2M^2}{\rho^2}}{\left(1-\frac{M}{\rho}\right)^5}\, ,
\end{eqnarray}

\begin{eqnarray}
D_8:=-\frac{5}{32}\left(\frac{M}{\rho}\right)^9\frac{1-\frac{2M}{\rho}+\frac{2M^2}{\rho^2}}{\left(1-\frac{M}{\rho}\right)^7}\, .
\end{eqnarray}

The way in which the $L^{(2)}$ operator couples modes is through the term proportional to $\,\sin^2\theta$. Specifically, when operating on a multipole moment $\ell$, the product  $\,\sin^2\theta\,P_{\ell}(\,\cos\theta)$ can be expanded to pure multipoles $\ell+2,\ell$, and $\ell-2$. We list useful identities in Appendix \ref{identities}. 

\section{Explicit treatment of a specific case}\label{full}

We now specialize to an explicit test case. We choose the initial multipole $\ell'=2$, and focus attention on the multipoles $\ell=0,2,4,6$. We list in Appendix \ref{app_full_equations} the evolution equations. Each equation has the form
\begin{equation}
\Box f^{(2n)}_{\ell}+V_{\ell}(\rho)\, f^{(2n)}_{\ell}=S^{(2n)}_{\ell}\, ,
\end{equation}
where the effective potential is given by 
\begin{equation}
V_{\ell}(\rho)=\frac{1}{\rho^2}\,\left(1-\frac{2M}{\rho}\right)\,\left[ \ell(\ell+1)+\frac{2M}{\rho}\right]\, ,
\end{equation}
and $S^{(2n)}_{\ell}$ is a sum of terms involving differential operators operating on lower--order iteration modes with multipoles $\ell-2,\ell$ and $\ell+2$. 
This set of inhomogeneous wave equations has the following property. Each inhomogeneous equation, like any other linear differential equation, has a solution which has the form of a general solution of the homogeneous equation plus a particular solution of the inhomogeneous equation. The only equation in the set that has a non-trivial, non-zero homogeneous solution is the one corresponding with the mode present in the initial data. In this case that is Eq.~(\ref{f02n}). Therefore, all excited modes (given by Eqs.~(\ref{f20n})--(\ref{f8_10n})) have as their solution only the particular solutions of the inhomogeneous differential equations. 

\begin{figure}
 \includegraphics[width=4.5cm]{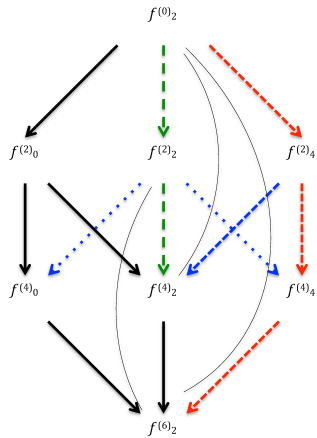}
\caption{The ten possible channels of excitations leading from the initial $f^{(0)}_2$ to the final $f^{(6)}_2$. The dominant channels (shown with solid arrows; see discussion below in Section \ref{dominant}) have one leg through $f^{(4)}_0$ that excites through $k=1$ (Channel (i)), and a second leg through $f^{(4)}_2$ that excites through $k=4(2),4(3)$ and $4(4)$ (Channel (ii)). The thin curves are excitations that couple directly modes separated by more than one order (Channels (viii), (ix), and (x)). In the dominant channel approximation we compute only the channels indicated by solid arrow, and neglect all other channels.}
\label{all_channels}
\end{figure}

There are ten possible channels of excitation from the initial $f^{(0)}_2$ mode to the final $f^{(6)}_2$ mode. The ten channels are as follows: (i) $f^{(0)}_2 \to f^{(2)}_0 \to f^{(4)}_0 \to f^{(6)}_2$; (ii) $f^{(0)}_2 \to f^{(2)}_0 \to f^{(4)}_2 \to f^{(6)}_2$; (iii) $f^{(0)}_2 \to f^{(2)}_2 \to f^{(4)}_2 \to f^{(6)}_2$; (iv) $f^{(0)}_2 \to f^{(2)}_2 \to f^{(4)}_0 \to f^{(6)}_2$; (v) $f^{(0)}_2 \to f^{(2)}_4 \to f^{(4)}_4 \to f^{(6)}_2$; (vi) $f^{(0)}_2 \to f^{(2)}_4 \to f^{(4)}_2 \to f^{(6)}_2$; (vii) $f^{(0)}_2 \to f^{(2)}_2 \to f^{(4)}_4 \to f^{(6)}_2$; (viii) $f^{(0)}_2  \to f^{(4)}_2 \to f^{(6)}_2$; (ix) $f^{(0)}_2 \to f^{(2)}_2 \to f^{(6)}_2$; and (x) $f^{(0)}_2  \to f^{(6)}_2$. The ten channels are shown schematically in Fig.~\ref{all_channels}. Notice that  channels (viii), (ix), and (x) include direct coupling between modes that are separated by more than one order. 

The numerical evolution scheme used in this paper is based on a characteristic grid that utilizes the $\rho_*$ (see Eq.~(\ref{rho_star})) and $t$ coordinates. Such a characteristic approach utilizes no boundary conditions. The code converges globally with second order. The grid separation that was used in this work take values given by, $\,\Delta\rho_* = \,\Delta t = 0.035$ horizon radii with the overall mass-scale set in practice as $M=0.5$ horizon radius.  The initial data for the zeroth order field in all studied cases, was chosen to be a smooth Gaussian distribution centered at $\rho_* = 0$ with a width of $8.0$ horizon radii. For all higher-order fields, the initial data were set equal to zero. All fields were evolved simultaneously using the same characteristic evolution scheme.  All $\ell'=2$ cases were performed in quadruple precision (128-bit). The $\ell'=4,6$ cases were performed in octal precision (256-bit).

\begin{figure}
 \includegraphics[width=8.0cm]{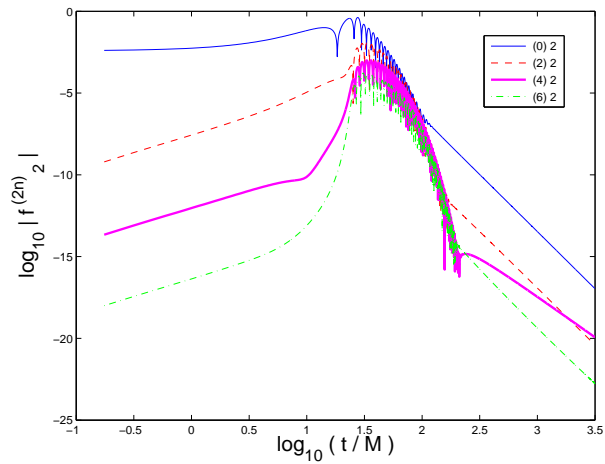}
\caption{The fields $f^{(2n)}_2$ for $(2n)=0$ (thin solid curve),2 (dashed curve), 4 (dash--dotted curve), and 6 (thick solid curve). At late times the mode corresponding to $(2n)=4$ dominates. The asymptotic decay rate of the $(2n)=6$ mode is the same as that of the $(2n)=4$, but this evolution is not long enough to show that. The eventual dominant mode is shown with a thicker curve. }
\label{f_2n_2}
\end{figure}

The quadrupole field $f_{2}$ is given to order $N$ by
\begin{equation}
f_{2}=\sum_{n=0}^N \left(\frac{a}{M}\right)^{2n} f_{2}^{(2n)}\, .
\end{equation}
The fields $f^{(2n)}_2$ are shown in Fig.~\ref{f_2n_2} for $(2n)=0,2,4,6$. Notice in Fig.~\ref{f_2n_2} that the full $\ell=2$ multipole decays at the same rate as the mode $f^{(0)}_2$ does, because the evolution is not long enough to show the dominance of the $f^{(4)}_2$ at very late times. It is only the slower decay rate of $f^{(4)}_2$ that shows that at very late times it would be the dominant mode. This is one of the benefits of the GPP approach: it allows us to find the ultimate dominant mode even when it is still subdominant. A (2+1)D approach would require much longer evolution times for the same initial data set to find the dominant mode, as the  $f_{2}$ field is dominated here until high values of the time parameter by its initial mode  $f_{2}^{(0)}$.

We write the late--time behavior of the mode $f^{(2n)}_{\ell}\sim t^{n}$. 
We find the decay rates\footnote{Notice that there is no confusion between the decay rate $n$ and the iteration order $(2n)$ as the latter is always written in brackets.} shown in Table \ref{table1}.

\begin{table}[htdp]
\caption{The asymptotic decay rates of the field modes with $\ell'=2$. At each expansion order in $a/M$ we list the asymptotic decay rate $n$ of each $\ell$ mode. In bold face we show the dominating order for each $\ell$ mode. In brackets we show  expected decay  rates we cannot resolve numerically with quad precision. }
\begin{center}
\begin{tabular}{||c||c|c|c|c|c|c||}
\hline
$(2n)\setminus\ell$  & 0 & 2 & 4 & 6 & 8 & 10 \\
 \hline
 \hline
0 & --  & -7 &  &   & &  \\
\hline
2& -{\bf 3} & -7 & -9 & &  & \\
\hline
4 & -3 & {\bf -5} & -9 & -11 & &  \\
\hline
6 & -3 & -5 & {\bf -7} & -11 & -13 &  \\
\hline
8 & -3 & -5 & (-7) & ({\bf -9}) & (-13) & (-15) \\
\hline
\hline
\end{tabular}
\end{center}
\label{table1}
\end{table}%

For each multipole $\ell$, the field is given by
\begin{equation}
\psi_{\ell}=\psi^{(0)}_{\ell}+\left(\frac{a}{M}\right)^2\psi^{(2)}_{\ell}+\left(\frac{a}{M}\right)^4\psi^{(4)}_{\ell}+\left(\frac{a}{M}\right)^6\psi^{(6)}_{\ell}+\left(\frac{a}{M}\right)^8\psi^{(8)}_{\ell}+\cdots
\end{equation}
so that clearly $\psi_{\ell}$ is dominated at late time by the slowest decaying $\psi^{(2n)}_{\ell}$. 
This table is fully consistent with the proposal of ZKB, that is it suggests that the dominant decay rate for the $\ell$ mode is given by $-n=\ell'+\ell+1$.

Specifically, in order to track the mode coupling mechanism, we separate the various source terms, and solve independently for each source term. The full solution is then just the sum of the solutions of the individual source terms for each equation. 
%\section*{Tracking the mode coupling mechanism}We propose to track the mode coupling mechanism in the following way: 
We choose to demonstrate the tracking of the excitation of the mode $f^{(6)}_2$ for two reasons: first,  because in Fig.~\ref{f_2n_2} we were unable to find its asymptotic decay rate, and second, because we expect it to have two dominant sources that contribute comparably. Specifically, both the source term that couples to $\ddot{f}^{(4)}_0$ and the source term that couples to $\,\Delta^2 f^{(4)}_2$ decay at late times like $t^{-5}$, and therefore will dominate at late times (all the other source terms decay faster). %(At a later step we could find out which is the dominant term in $\,\Delta^2$.) Which of these two sources dominates depends therefore on their amplitudes, as they both have the same late time decay rate. 

\begin{figure}
 \includegraphics[width=8.0cm]{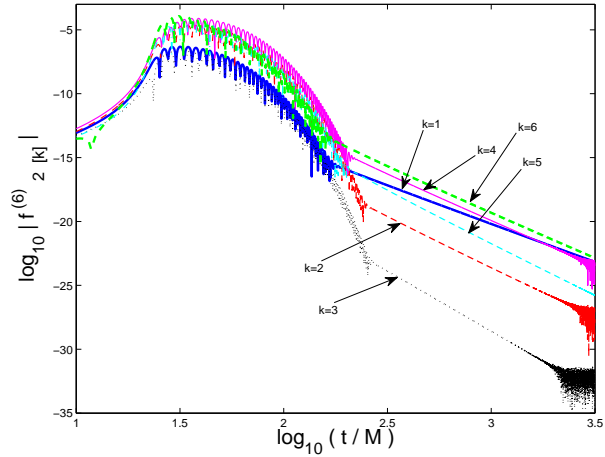}
\caption{The partial fields $f^{(6)}_{2\,[k]}$ as functions of time. Shown are the cases $k=1$ (thick solid curve), $k=2$ (dashed curve), $k=3$ (dotted curve), $k=4$ (thin solid curve), $k=5$ (dashed curve), and $k=6$ (thick dashed curve),}
\label{f62k}
\end{figure}

We therefore split the equation for the $f^{(6)}_2$ mode into 7 separate equations, one homogeneous equation and 6 inhomogeneous equations, each one with a single source term. The homogeneous equation must have an identically zero solution (for the vanishing initial data for modes that are not at (0) order), and therefore the full solution of each equation is the particular solution of the inhomogeneous equation. The sum of the 6 particle solutions should equal the solution of the full equation for the $f^{(6)}_2$ mode. Specifically, we write 
\begin{equation}
f^{(6)}_{2}=\sum_{k=0}^{6}f^{(6)}_{2\;[k]}\, .
\end{equation}
where $f^{(6)}_{2\;[0]}=0$ identically. The equations of motion for the partial fields $f^{(6)}_{2\;[k]}$ of the mode $f^{(6)}_{2}$ are listed explicitly in Appendix \ref{app_partial_fields}.  The partial fields $f^{(6)}_{2\;[k]}$ are shown in Fig.~\ref{f62k}. Notice that while clearly the partial field $f^{(6)}_{2\;[1]}$ dominates at late times, the evolved tail portion of the field is dominated by the $f^{(6)}_{2\;[6]}$ partial field. Longer evolution is needed in order to have the mode $f^{(6)}_{2}$ dominated by the partial field $f^{(6)}_{2\;[1]}$. The evolution shown, which may be naively thought of as sufficiently long, only shows an intermediate behavior if data are collected only for the full mode $f^{(6)}_{2}$. The benefit of the GPP approach is that it allows us to recognize a subdominant partial field, that would become dominant at asymptotic times. Notice also that the partial field $f^{(6)}_{2\,[4]}$ in itself is not yet asymptotic, but rather is shown in a regime which is intermediate. We discuss the partial field $f^{(6)}_{2\,[4]}$ in greater detail below.

\begin{figure}
 \includegraphics[width=8.0cm]{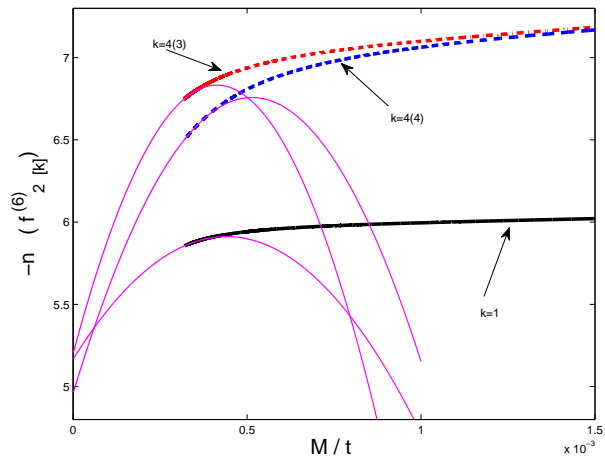}
\caption{The local power indices for $f^{(6)}_{2\,[1]}$ (thick solid curve), $f^{(6)}_{2\,[4(3)]}$ (dash--dotted curve), and $f^{(6)}_{2\,[4(4)]}$ (dashed curve). Thick curves are the local power indices. The thin  curves are quadratic extrapolations to asymptotically late time.}
\label{f62_n}
\end{figure}

\begin{figure}
 \includegraphics[width=8.0cm]{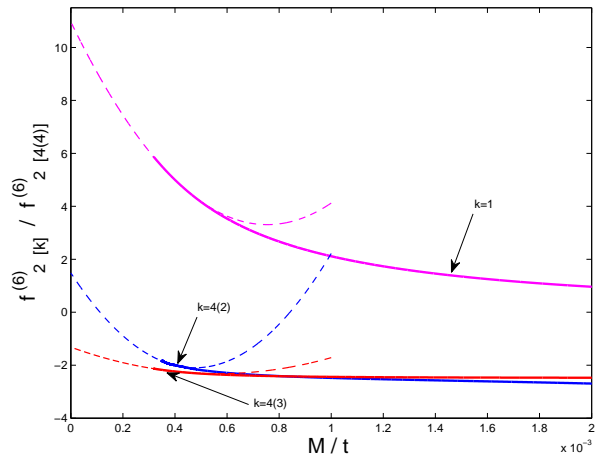}
\caption{The ratios $f^{(6)}_{2\,[1]}/f^{(6)}_{2\,[4(4)]}$, $f^{(6)}_{2\,[4(2)]}/f^{(6)}_{2\,[4(4)]}$, and $f^{(6)}_{2\,[4(3)]}/f^{(6)}_{2\,[4(4)]}$. Thick curves are the computed ratios. The thin  curves are quadratic extrapolations to asymptotically late time.}
\label{f62_ratio}
\end{figure}

We find that $\Delta_{(2)}^2 f^{(4)}_2,\Delta_{(3)}^2f^{(4)}_2$ and $\Delta_{(4)}^2f^{(4)}_2$ (which are defined in Appendix \ref{app_partial_fields}) decay asymptotically in time at the same rate (see Figs.~\ref{f62_n},\ref{f62_ratio}). When the partial wave arising from Eq.~(\ref{f62n4}) dominates at late time, all three need to be considered.     
Figure \ref{f62_n} does not include information about the case $k=4(2)$, because it is too noisy due to the second spatial derivative in the source. We present in Fig.~\ref{f62_ratio} the result that the $k=4(2)$ case decays with the same rate as the three cases shown in Fig.~\ref{f62_n} by plotting the ratios $f^{(6)}_{2\;[k]}/f^{(6)}_{2\;[(4(4)]}$ for $k=1,4(2)$ and $4(3)$. When this ratio approaches a non-zero constant as $M/t\to 0$, the two decay rates are asymptotically equal, and the limiting ratio is the ratio of amplitudes. We therefore infer that the late--time partial field $f^{(6)}_2$ is dominated by $k=1$, and that the contributions of $k=4(2),4(3)$ and $4(4)$ are an order of magnitude smaller than the contribution of $k=1$. However, to find the full partial field, one needs to include those contributions too. 

In Table \ref{table2} we show the asymptotic decay rates of the partial fields, and also the decay rates of the respective source terms in Eqs.~(\ref{f62n1})--(\ref{f62n6}) and Eqs.~(\ref{f62n4_1})--(\ref{f62n4_4}). We find correspondence of the decay rate of the partial field and the decay rate of the source term. As expected for linear differential equations, the same decay rate of the source term leads to the same decay rate of the partial field. Notably, the field in the source term that couples to the $k=1$ case is different from the field in the source term that couples to the $k=4(j)$ cases. Therefore, the dominant channel of excitation bifurcates, and exhibits a more complex structure than that proposed by GPP. (One may also say that there are two dominant channels with the same asymptotic decay rates.) Figure \ref{all_channels} shows the dominant channel of mode excitation for the mode $f^{(6)}_2$. Although the modes $f^{(4)}_2$ and $f^{(4)}_0$ decay asymptotically at different rates, they contribute to partial fields of $f^{(6)}_2$ that decay asymptotically at equal rates. The reason of this behavior is that the differential operator for the case $k=1$ is a second-order temporal differential operator, whereas the differential operator for the case $k=4$ (more specifically, the cases $k=4(2),4(3),4(4)$) includes terms with no temporal differentiation. 

Based on our reasoning, we predict that a similar analysis of the mode $f^{(4)}_2$ would produce results analogous to those appearing in Table \ref{table2}. These predicted values appear in Table \ref{table3}. We were able to check only in part the entries in Table \ref{table3} due to the high level of numerical noise.

\begin{table}[htdp]
\caption{The asymptotic decay rates of the different terms contributing to $f^{(6)}_2$. 
The index $n$ corresponds to the 
decay rate of the $f^{(6)}_{2\,[k]}$ partial field. The index $m$ corresponds to the decay rate of the source term $S^{(6)}_{2\,[k]}$.}
\begin{center}
\begin{tabular}{||c||c|c|c|c|c|c|c|c|c||}
\hline
$k$  & 1 & 2 & 3 & 4(1) & 4(2) & 4(3) & 4(4) & 5 & 6 \\
 \hline
$n$ & -5  &  -7 & -9 & -7  & -5 & -5 & -5 & -7 & -7  \\
 \hline
$m$ & -5  &  -7 & -11 & -7  & -5 & -5 & -5 & -9 & -9  \\
\hline
\hline
\end{tabular}
\end{center}
\label{table2}
\end{table}%

\begin{table}[htdp]
\caption{The asymptotic decay rates of the different terms contributing to $f^{(4)}_2$. 
The index $n$ corresponds to the 
decay rate of the $f^{(4)}_{2\,[k]}$ partial field. The index $m$ corresponds to the decay rate of the source term $S^{(4)}_{2\,[k]}$. In brackets we show 
estimated values. The combination of all the terms in brackets decay with rate of $-7$. }
\begin{center}
\begin{tabular}{||c||c|c|c|c|c|c|c|c|c||}
\hline
$k$  & 1 & 2 & 3 & 4(1) & 4(2) & 4(3) & 4(4) & 5  \\
 \hline
$n$ & -5  &  -7 & -9 & (-9)  & (-7) & (-7) & (-7) & -7   \\
 \hline
$m$ & -5  &  -9 & -11 & -9  & -7 & -7 & -7 & -9  \\
\hline
\hline
\end{tabular}
\end{center}
\label{table3}
\end{table}%

\section{Specializing to the dominant channel}\label{dominant}

Our discussion in the preceding Section allows us to identify the dominant channel given the initial multipole $\ell'$ and the multipole of interest $\ell$. Evolving the field equations for the dominant channel only is much simpler, as there are fewer equations, and the structure of the equations is simpler. On the other hand, by analyzing the dominant channel only, one may find the correct decay rate at asymptotically late time, but only an approximation for the amplitude of the field. Below, we show that this approximation is improving fast at very late times. 

We illustrate the method to identify the dominant channel with an explicit example. Consider $\ell'=2$ and $\ell=2$. In this case we showed above
that the asymptotic decay rate is $-n=5$, and that this rate is first achieved at order $(2n)=4$. At zeroth ($(2n)=0$) and first ($(2n)=2$) orders we fnd that the decay rates are given by $-n^{(0)}_2=7$, $-n^{(2)}_0=3$, $-n^{(2)}_2=7$, and $-n^{(2)}_4=9$. When examining the equation for the mode of interest, Eq.~(\ref{f42n}) for the mode $f^{(4)}_2$, we notice there are five source terms. The source terms decay at different rates at late times, so that we may evaluate which one dominates. Specifically, $f^{(2)}_{0\; ,tt}\sim t^{-5}$, $f^{(2)}_{2\; ,tt}\sim t^{-9}$, $f^{(2)}_{4\; ,tt}\sim t^{-11}$, $\,\Delta^2 f^{(2)}_{2}\sim t^{-7}$, and $f^{(0)}_{2\; ,tt}\sim t^{-9}$. Clearly, at late times the mode $f^{(4)}_2$ is dominated by the first source term, that is by the $f^{(2)}_0$ mode. Therefore, the dominant channel for this case is $f^{(0)}_2 \to f^{(2)}_0 \to f^{(4)}_2$. Similar considerations allow us to identify the dominant channel in all other cases too. We comment that at earlier times other channels may overwhelm the asymptotically dominant channel. The latter dominates only at very late times. 

We focus attention on three test cases: first, the case $\ell'=2$, $\ell=0,2,4,6$, second $\ell'=4$, $\ell=4$, and third, the case $\ell'=6$, $\ell=0,2,4,6$.

\begin{table}[htdp]
\caption{The asymptotic decay rates of the fields of the dominant channels described in Sections \ref{ss3} (curly brackets), \ref{ss1} (square brackets) and \ref{ss2} (no brackets). In round brackets we show  expected decay  rates we cannot obtain due to numerical errors. }
\begin{center}
\begin{tabular}{||c||c|c|c|c|c|c||}
\hline
$(2n)\setminus\ell$  & 0 & 2 & 4 & 6  \\
 \hline
 \hline
0 &  &\{7\}  & [11] & 15   \\
\hline
2 &\{3\}  & [7] & 11 &  \\
\hline
4 & [5] & 9,\{5\} &  &   \\
\hline
6 & 7 & [7] & \{7\} &  \\
\hline
8 &  & 9 & [9]  & \{9\} \\
\hline
10 & & & (11) & \\
\hline
12 & & & & (13) \\
\hline
\hline
\end{tabular}
\end{center}
\label{table4}
\end{table}%

\subsection{The case $\ell'=2$, $\ell=0,2,4,6$}\label{ss3}

Our first case is $\ell'=2$, and $\ell=2,4$ or $6$. For the channel of interest we take

\begin{eqnarray}
\psi^{(0)}&=&\frac{1}{\rho}f^{(0)}_2(t,\rho)P_2(\,\cos\theta)\\
\psi^{(2)}&=&\frac{1}{\rho}f^{(2)}_0(t,\rho)P_0(\,\cos\theta)\\
\psi^{(4)}&=&\frac{1}{\rho}f^{(4)}_2(t,\rho)P_2(\,\cos\theta)\\
\psi^{(6)}&=&\frac{1}{\rho}f^{(6)}_4(t,\rho)P_4(\,\cos\theta)\\
\psi^{(8)}&=&\frac{1}{\rho}f^{(8)}_6(t,\rho)P_6(\,\cos\theta)\, .
\end{eqnarray}
The field equations are listed in Appendix \ref{app:26}. 

This case is of particular interest, because it allows us to compare the accuracy of the dominant channel approximation with the full set of equations studied in Section \ref{full}. Naturally, the fields $\psi^{(0)}_2$ and $\psi^{(2)}_0$ are each identical between the full and approximate calculations, because their respective field equations are exactly the same. We do not expect a good agreement between the dominant branch fields and the full fields at early times. Indeed, the essence of the approximation is to evaluate the dominant contribution to the source term based on the asymptotic behavior, and discarding the source terms that are subdominant asymptotically. As is clearly the case, asymptotically subdominant fields may be important at earlier times. 

\begin{figure}
 \includegraphics[width=8.0cm]{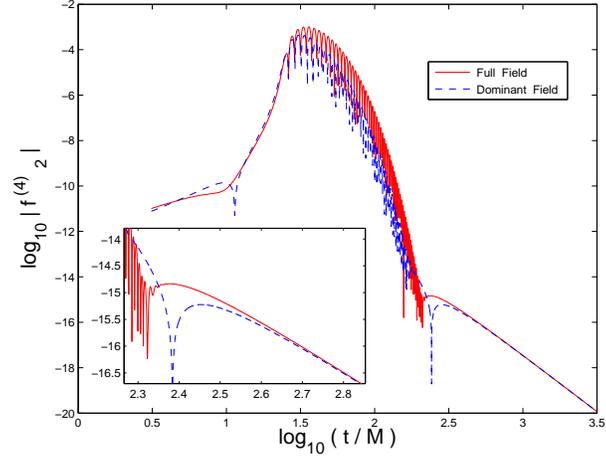}
\caption{The field $\psi^{(4)}_2$ as a function of time for the full field (solid) and the dominant channel approximation (dashed).}
\label{fig42}
\end{figure}

\begin{figure}
 \includegraphics[width=8.0cm]{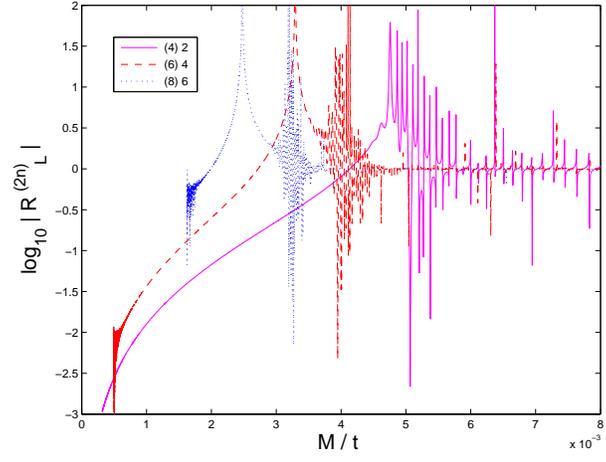}
\caption{The relative difference $R^{(2n)}_{\ell}$ as a function of inverse time for $\psi^{(4)}_2$ (solid) , $\psi^{(6)}_4$ (dashed), and $\psi^{(8)}_6$ (dotted).}
\label{R}
\end{figure}

\begin{figure}
 \includegraphics[width=8.0cm]{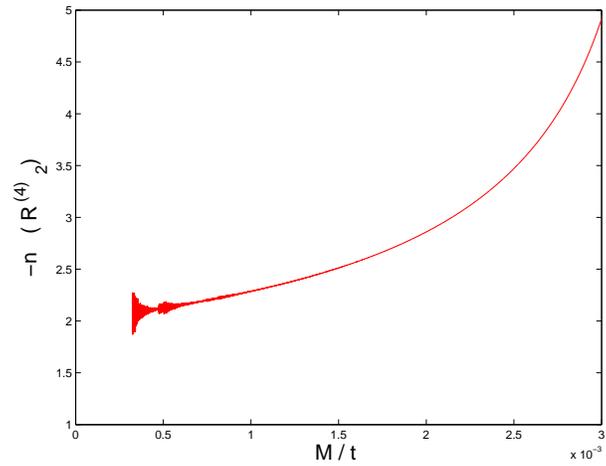}
\caption{The rate with which $R^{(4)}_{2}$ decays to zero at late times, $R^{(4)}_{2}\sim t^{n}$.}
\label{R_conv}
\end{figure}

In Fig.~\ref{fig42} we show the $\psi^{(4)}_2$ field from the full computation of Section \ref{full} and from the dominant channel approach we are using here. Clearly, they both have the same asymptotic decay rate. Indeed the two fields are very different at early times, but converge at late times. To quantify the agreement we define the relative difference of the two results, specifically, $R^{(2n)}_{\ell}:=1-\psi^{(2n)}_{\ell, \; {\rm dominant}}/\psi^{(2n)}_{\ell, \;  {\rm full}}$, and plot in Fig.~\ref{R} $R^{(2n)}_{\ell}$ as a function of (inverse) time for the three non--trivial fields. We find that at late times the difference between the two fields, $R^{(2n)}_{\ell}$, approaches zero like $t^{-2}$ (see Fig.~\ref{R_conv}). We infer that at very late time indeed the dominant channel approach is very useful not just in finding the decay rate of the field of interest, but also in finding its amplitude. 

\subsection{The case $\ell'=4$, $\ell=4$}\label{ss1}

We are looking for the case $\ell'=4$, $\ell=4$. The dominant channel of excitations includes the following fields. 
\begin{eqnarray}
\psi^{(0)}&=&\frac{1}{\rho}f^{(0)}_4(t,\rho)P_4(\,\cos\theta)\\
\psi^{(2)}&=&\frac{1}{\rho}f^{(2)}_2(t,\rho)P_2(\,\cos\theta)\\
\psi^{(4)}&=&\frac{1}{\rho}f^{(4)}_0(t,\rho)P_0(\,\cos\theta)\\
\psi^{(6)}&=&\frac{1}{\rho}f^{(6)}_2(t,\rho)P_2(\,\cos\theta)\\
\psi^{(8)}&=&\frac{1}{\rho}f^{(8)}_4(t,\rho)P_4(\,\cos\theta)\, .
\end{eqnarray}
The respective equations governing these fields are listed in Appendix \ref{app:44}. The fields are shown in Fig.~\ref{f44fields}. This case was already studies in (2+1)D by ZKB, who found $-n=9$ for the mode of interest. Our GPP approach allows us to track how this asymptotic decay rate of $-n=9$ comes about. The initial data with multipole $\ell'=4$, or zeroth order field $\psi^{(0)}$ decays asymptotically as $-n=2\times 4+3=11$. At order (2n)=2 the field $\psi^{(2)}$ decays asymptotically as $-n=4+2+1=7$, at order (2n)=4 the field $\psi^{(4)}$ decays asymptotically as $-n=4+0+1=5$, at order (2n)=6 the field $\psi^{(6)}$ decays asymptotically as $-n=4+2+1=7$, and at order (2n)=8 the field $\psi^{(8)}$ decays asymptotically as $-n=4+4+1=9$. Both the fields $\psi^{(0)}$ and $\psi^{(8)}$ are multipoles with $\ell=4$, but the latter decays slower at late times, and therefore will asymptotically dominate, even though it is many orders of magnitude smaller in amplitude than the former over the length of our numerical simulation, as shown in Fig.~\ref{f44fields}. In Fig.~\ref{f44} we show the local decay rates of the five modes computed in the dominant channel. Again, the GPP approach allows us to determine the asymptotic decay rate even though the mode in question is subdominant. If one were to solve for the full $\ell=4$ multipole, the required evolution would have to be much longer to find the asymptotic behavior instead of an intermediate behavior.

\begin{figure}
 \includegraphics[width=8.0cm]{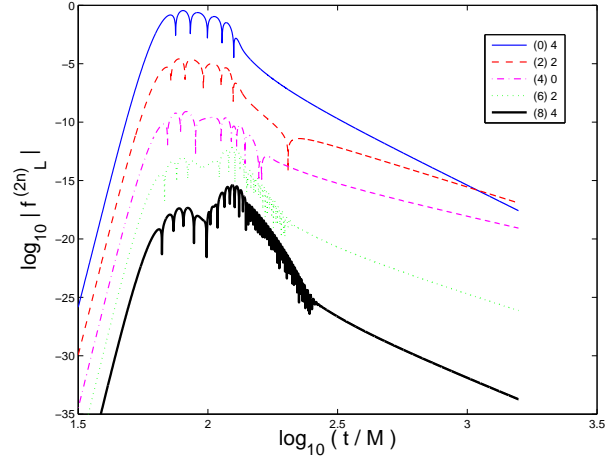}
\caption{The fields appearing in the dominant branch for the case $\ell'=4$, $\ell=4$ as function of the time. Shown are $f^{(0)}_4$ (thin solid curve), $f^{(2)}_2$ (dashed curve), $f^{(4)}_0$ (dash--dotted curve), $f^{(6)}_2$ (dotted curve), and $f^{(8)}_4$ (thick solid curve). }
\label{f44fields}
\end{figure}

\begin{figure}
 \includegraphics[width=8.0cm]{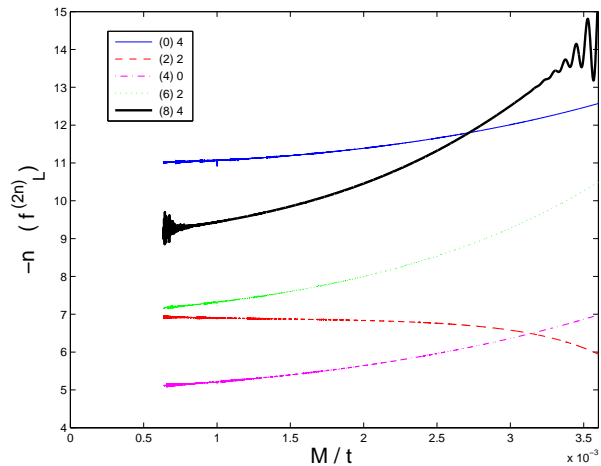}
\caption{The local decay rates $-n$ for the  fields appearing in the dominant branch for the case $\ell'=4$, $\ell=4$. Shown are the decay rates for the fields $f^{(0)}_4$ (thin solid curve), $f^{(2)}_2$ (dashed curve), $f^{(4)}_0$ (dash--dotted curve), $f^{(6)}_2$ (dotted curve), and $f^{(8)}_4$ (thick solid curve). }
\label{f44}
\end{figure}

\subsection{The case $\ell'=6$, $\ell=0,2,4,6$}\label{ss2}

Our next case is $\ell'=6$, and $\ell=0,2,4$ or $\ell=6$. For the channel of interest we take
\begin{eqnarray}
\psi^{(0)}&=&\frac{1}{\rho}f^{(0)}_6(t,\rho)P_6(\,\cos\theta)\\
\psi^{(2)}&=&\frac{1}{\rho}f^{(2)}_4(t,\rho)P_4(\,\cos\theta)\\
\psi^{(4)}&=&\frac{1}{\rho}f^{(4)}_2(t,\rho)P_2(\,\cos\theta)\\
\psi^{(6)}&=&\frac{1}{\rho}f^{(6)}_0(t,\rho)P_0(\,\cos\theta)\\
\psi^{(8)}&=&\frac{1}{\rho}f^{(8)}_2(t,\rho)P_2(\,\cos\theta)\\
\psi^{(10)}&=&\frac{1}{\rho}f^{(10)}_4(t,\rho)P_4(\,\cos\theta)\\
\psi^{(12)}&=&\frac{1}{\rho}f^{(12)}_6(t,\rho)P_6(\,\cos\theta)\, .
\end{eqnarray}
The field equations are listed in Appendix \ref{app:66}. The fields are shown in Fig.~\ref{f66fields} and the corresponding local decay rates are shown in Fig.~\ref{f66}. The small amplitude of high--order excitations is shown here in a sharp way: numerical noise is obtained early for the $f^{(10)}_4$ field and even earlier for the $f^{(12)}_6$ field, so that we cannot determine with much accuracy the respective asymptotic decay rates with our  computation. The decay rates for the fields we can compute are all in agreement with the proposed formula: the mode the includes the initial data decays at late times as $-n=2\ell'+3$, and all other excited modes decay as $-n=\ell'+\ell+1$. 

\begin{figure}
 \includegraphics[width=8.0cm]{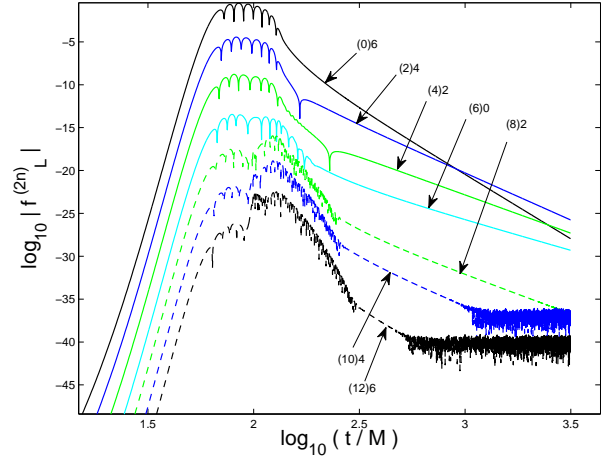}
\caption{The fields appearing in the dominant branch for the case $\ell'=6$, $\ell=0,2,4,6$. The orders $(0),(2),(4)$ and $(6)$ are shown with solid curves, and the orders $(8)$, $(10)$, and $(12)$ are shown with dashed curves. }
\label{f66fields}
\end{figure}

\begin{figure}
 \includegraphics[width=8.0cm]{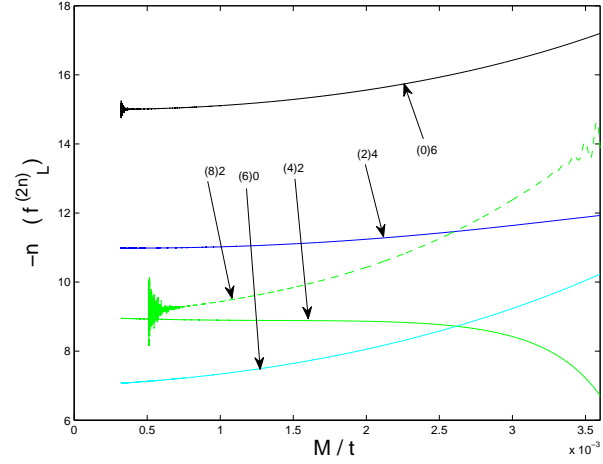}
\caption{The local decay rates $-n$ for the  fields appearing in the dominant branch for the case $\ell'=6$, $\ell=0,2,4,6$. The orders $(0),(2),(4)$ and $(6)$ are shown with solid curves, and the order $(8)$ is shown with dashed curves.}
\label{f66}
\end{figure}

\section*{Acknowledgments}

The authors are indebted to Jorge Pullin and to Richard Price for useful and stimulating discussions. L.M.B.~acknowledges research support from NSF Grant Nos.~PHY-1249302 and DUE-1300717. G.K.~acknowledges research support from NSF Grant Nos.~PHY-1016906, CNS-0959382, PHY-1135664 and PHY-1303724, and from the US Air Force Grant Nos.~FA9550-10-1-0354 and 10-RI-CRADA-09.

\begin{appendix}

\section{Properties of the $\rho$ coordinate}\label{app_a}

The radial coordinate $\rho$ is defined in Eq.~(\ref{rho_def}) as 
\begin{equation}
r:=M+\sqrt{\rho^2-2M\rho+M^2-a^2}\, .
\end{equation} The following relations prove to be useful:

\begin{equation}\label{rho_star}
\frac{\partial}{\,\partial \rho} := \frac{1}{1-\frac{2M}{\rho}} \frac{\partial}{\,\partial \rho_*}
\end{equation}
(which defines the `tortoise' coordinate, $\rho_*=\rho+2M\,\log \left(\frac{\rho}{2M}-1\right)\, .$

\begin{equation}
\frac{\partial^2}{\,\partial \rho^2} = \frac{1}{\left(1-\frac{2M}{\rho}\right)^2} \left(\frac{\partial^2}{\,\partial {\rho_*}^2}-\frac{2M}{\rho^2} 
\frac{\partial}{\,\partial \rho^*}\right) 
\end{equation}

\begin{equation}
\frac{\,d\rho}{\,dr}=\frac{\sqrt{\rho^2-2M\rho+M^2-a^2}}{\rho-M}=\frac{r-M}{\rho-M}
\end{equation}

\begin{equation}
\frac{\,d^2\rho}{\,dr^2}=\frac{a^2}{(\rho-M)^3}
\end{equation}

\begin{equation}
\frac{\,\partial}{\,\partial r}=\frac{\sqrt{\rho^2-2M\rho+M^2-a^2}}{\rho-M}\,\frac{\,\partial}{\,\partial\rho}
\end{equation}

\begin{equation}
\frac{\,\partial^2}{\,\partial r^2}=\frac{\rho^2-2M\rho+M^2-a^2}{(\rho-M)^2}\,\frac{\,\partial^2}{\,\partial\rho^2}+\frac{a^2}{(\rho-M)^3}\,\frac{\,\partial}{\,\partial\rho}
\end{equation}

\begin{equation}
\,\partial_{\rho_*}\left[ \frac{1}{\rho}f\right] = \frac{1}{\rho}\,\partial_{\rho_*}f-\frac{1}{\rho^2}\left(1-\frac{2M}{\rho}\right)f
\end{equation}

\begin{equation}
\,\partial^2_{\rho_*}\left[ \frac{1}{\rho}f\right] = \frac{1}{\rho}\,\partial^2_{\rho_*}f-\frac{2}{\rho^2}\left(1-\frac{2M}{\rho}\right)\,\partial_{\rho_*}f
+\frac{2}{\rho^3}\left(1-\frac{2M}{\rho}\right)\left(1-\frac{3M}{\rho}\right)f\, .
\end{equation}

\section{The differential operators}\label{app_operators}

The differential operators that appear in Eq.~(\ref{iterative_equation}) are given by:
\begin{eqnarray}
L^{(0)}[\phi]&=&\frac{\rho^3}{\rho-2M}\frac{\,\partial^2\phi}{\,\partial t^2}-\rho(\rho-2M)\frac{\,\partial^2\phi}{\,\partial \rho^2}-2(\rho-M)\frac{\,\partial\phi}{\,\partial \rho}\nonumber \\
&-&\frac{1}{\,\sin\theta}\frac{\,\partial}{\,\partial\theta}\left( \,\sin\theta\frac{\,\partial\phi}{\,\partial\theta} \right)\\
&=&\frac{\rho^2}{1-\frac{2M}{\rho}}\left(\frac{\,\partial^2\phi}{\,\partial t^2}-\frac{\,\partial^2\phi}{\,\partial \rho_*^2}\right)-2\rho\frac{\,\partial\phi}{\,\partial \rho_*}
-\frac{1}{\,\sin\theta}\frac{\,\partial}{\,\partial\theta}\left( \,\sin\theta\frac{\,\partial\phi}{\,\partial\theta} \right)
\\
L^{(2)}[\phi]&=&M^2\,\left[-\left(\,\frac{2M}{\rho}\frac{1}{1-\frac{M}{\rho}}\frac{1}{1-\frac{2M}{\rho}}+ \,\sin^2\theta\right)\frac{\,\partial^2\phi}{\,\partial t^2}\right.  \nonumber \\
&+& \left. \frac{1-\frac{2M}{\rho}}{\left(1-\frac{M}{\rho}\right)^2}
\frac{\,\partial^2\phi}{\,\partial \rho^2}+\frac{1}{\rho}\,\frac{1-\frac{2M}{\rho}+\frac{2M^2}{\rho^2}}{\left(1-\frac{M}{\rho}\right)^3}\,\frac{\,\partial\phi}{\,\partial\rho}
\right]\\
&=&M^2\,\left[-\left(\,\frac{2M}{\rho}\frac{1}{1-\frac{M}{\rho}}\frac{1}{1-\frac{2M}{\rho}}+ \,\sin^2\theta\right)\frac{\,\partial^2\phi}{\,\partial t^2}\right.  \nonumber \\
&+&\left.\frac{1}{\left(1-\frac{2M}{\rho}\right)\left(1-\frac{M}{\rho}\right)^2}\frac{\,\partial^2\phi}{\,\partial \rho_*^2}+\frac{1}{\rho}\,\frac{1-\frac{2M}{\rho}}{\left(1-\frac{M}{\rho}\right)^3}
\frac{\,\partial\phi}{\,\partial \rho_*}\right]\\
L^{(4)}[\phi]&=&-\frac{1}{2}\frac{M^5}{\rho^3}\frac{1}{\left(1-\frac{M}{\rho}\right)^3}\frac{1-\frac{2M}{\rho}+\frac{2M^2}{\rho^2}}{1-\frac{2M}{\rho}}
\frac{\,\partial^2\phi}{\,\partial t^2}\\
L^{(6)}[\phi]&=&-\frac{1}{4}\frac{M^7}{\rho^5}\frac{1}{\left(1-\frac{M}{\rho}\right)^5}\frac{1-\frac{2M}{\rho}+\frac{2M^2}{\rho^2}}{1-\frac{2M}{\rho}}
\frac{\,\partial^2\phi}{\,\partial t^2}\\
L^{(8)}[\phi]&=&-\frac{5}{32}\frac{M^9}{\rho^7}\frac{1}{\left(1-\frac{M}{\rho}\right)^7}\frac{1-\frac{2M}{\rho}+\frac{2M^2}{\rho^2}}{1-\frac{2M}{\rho}}
\frac{\,\partial^2\phi}{\,\partial t^2}\\
L^{(10)}[\phi]&=&-\frac{7}{64}\frac{M^{11}}{\rho^9}\frac{1}{\left(1-\frac{M}{\rho}\right)^9}\frac{1-\frac{2M}{\rho}+\frac{2M^2}{\rho^2}}{1-\frac{2M}{\rho}}
\frac{\,\partial^2\phi}{\,\partial t^2}\\
L^{(12)}[\phi]&=&-\frac{21}{256}\frac{M^{13}}{\rho^{11}}\frac{1}{\left(1-\frac{M}{\rho}\right)^{11}}\frac{1-\frac{2M}{\rho}+\frac{2M^2}{\rho^2}}{1-\frac{2M}{\rho}}
\frac{\,\partial^2\phi}{\,\partial t^2}
\end{eqnarray}

\section{Useful identities}\label{identities}

In construction of the field equations, the following identities are useful. These are the identities that govern the mode coupling. 

\begin{equation}
\,\sin^2\theta \;P_0(\,\cos\theta)= -\frac{2}{3}\;P_2(\,\cos\theta)+\frac{2}{3}\;P_0(\,\cos\theta)
\end{equation}

\begin{equation}
\,\sin^2\theta \;P_2(\,\cos\theta)=-\frac{12}{35}\;P_4(\,\cos\theta)+\frac{10}{21}\;P_2(\,\cos\theta)-\frac{2}{15}\;P_0(\,\cos\theta)
\end{equation}

\begin{equation}
\sin^2\theta \; P_4(\cos\theta) =
-\frac{10}{33}\; P_6(\cos\theta)
+ \frac{38}{77}\; P_4(\cos\theta)
-\frac{4}{21}\; P_2(\cos\theta)
\end{equation}

\begin{equation}
\sin^2\theta \; P_6(\cos\theta) =
-\frac{56}{195}\; P_8(\cos\theta)
+ \frac{82}{165}\; P_6(\cos\theta)
-\frac{30}{143}\; P_4(\cos\theta)
\end{equation}

\begin{equation}
\sin^2\theta \; P_8(\cos\theta) =
-\frac{90}{323}\; P_{10}(\cos\theta)
+ \frac{142}{285}\; P_8(\cos\theta)
-\frac{56}{255}\; P_6(\cos\theta)
\end{equation}

\section{The Evolution Equations for $\ell'=2$}\label{app_full_equations}
The full equations for the modes for $\ell'=2$ are given by

\begin{eqnarray}\label{f02n}
\,\Box f^{(0)}_{2}+\frac{1}{\rho^2}\,\left(1-\frac{2M}{\rho}\right)\,\left(6+\frac{2M}{\rho}\right)f^{(0)}_2=0 \end{eqnarray}

\begin{eqnarray}\label{f20n}
\,\Box f^{(2)}_{0}+\frac{1}{\rho^2}\,\left(1-\frac{2M}{\rho}\right)\,\left(\frac{2M}{\rho}\right)f^{(2)}_0=-\frac{2}{15}\left(\frac{M}{\rho}\right)^2\left(1-\frac{2M}{\rho}\right)f^{(0)}_{2\; ,tt} \end{eqnarray}

\begin{eqnarray}\label{f22n}
\,\Box f^{(2)}_{2}+\frac{1}{\rho^2}\,\left(1-\frac{2M}{\rho}\right)\,\left(6+\frac{2M}{\rho}\right)f^{(2)}_2&=&\frac{10}{21}\left(\frac{M}{\rho}\right)^2\left(1-\frac{2M}{\rho}\right)f^{(0)}_{2\; ,tt} \nonumber \\
&-&\left(\frac{M}{\rho}\right)^2\,\Delta^2 f^{(0)}_{2}
\end{eqnarray}

\begin{eqnarray}\label{f24n}
\,\Box f^{(2)}_{4}+\frac{1}{\rho^2}\,\left(1-\frac{2M}{\rho}\right)\,\left(20+\frac{2M}{\rho}\right)f^{(2)}_4=-\frac{12}{35}\left(\frac{M}{\rho}\right)^2\left(1-\frac{2M}{\rho}\right)f^{(0)}_{2\; ,tt} \end{eqnarray}

\begin{eqnarray}\label{f40n}
\,\Box f^{(4)}_{0}+\frac{1}{\rho^2}\,\left(1-\frac{2M}{\rho}\right)\,\left(\frac{2M}{\rho}\right)f^{(4)}_0&=&\frac{2}{3}\left(\frac{M}{\rho}\right)^2\left(1-\frac{2M}{\rho}\right)f^{(2)}_{0\; ,tt} \nonumber \\
&-& \frac{2}{15}\left(\frac{M}{\rho}\right)^2\left(1-\frac{2M}{\rho}\right)f^{(2)}_{2\; ,tt} \nonumber \\
&-&\left(\frac{M}{\rho}\right)^2\,\Delta^2f^{(2)}_0
\end{eqnarray}

\begin{eqnarray}\label{f42n}
\,\Box f^{(4)}_{2}+\frac{1}{\rho^2}\,\left(1-\frac{2M}{\rho}\right)\,\left(6+\frac{2M}{\rho}\right)f^{(4)}_2&=&- \frac{2}{3}\left(\frac{M}{\rho}\right)^2\left(1-\frac{2M}{\rho}\right)f^{(2)}_{0\; ,tt} \nonumber \\
&+& \frac{10}{21}\left(\frac{M}{\rho}\right)^2\left(1-\frac{2M}{\rho}\right)f^{(2)}_{2\; ,tt} \nonumber \\
&-&\frac{4}{21}\left(\frac{M}{\rho}\right)^2\left(1-\frac{2M}{\rho}\right)f^{(2)}_{4\; ,tt} \nonumber \\
&-&\left(\frac{M}{\rho}\right)^2\,\Delta^2 f^{(2)}_2\nonumber \\
&-&D_4\, f^{(0)}_{2\; ,tt}
\end{eqnarray}

\begin{eqnarray}\label{f44n}
\,\Box f^{(4)}_{4}+\frac{1}{\rho^2}\,\left(1-\frac{2M}{\rho}\right)\,\left(20+\frac{2M}{\rho}\right)f^{(4)}_4&=&- \frac{12}{35}\left(\frac{M}{\rho}\right)^2\left(1-\frac{2M}{\rho}\right)f^{(2)}_{2\; ,tt} \nonumber \\
&+& \frac{38}{77}\left(\frac{M}{\rho}\right)^2\left(1-\frac{2M}{\rho}\right)f^{(2)}_{4\; ,tt} \nonumber \\
&-&\left(\frac{M}{\rho}\right)^2\,\Delta^2 f^{(2)}_4
\end{eqnarray}

\begin{eqnarray}\label{f46n}
\,\Box f^{(4)}_{6}+\frac{1}{\rho^2}\,\left(1-\frac{2M}{\rho}\right)\,\left(42+\frac{2M}{\rho}\right)f^{(4)}_6=- \frac{10}{33}\left(\frac{M}{\rho}\right)^2\left(1-\frac{2M}{\rho}\right)f^{(2)}_{4\; ,tt} 
\end{eqnarray}

\begin{eqnarray}\label{f60n}
\,\Box f^{(6)}_{0}+\frac{1}{\rho^2}\,\left(1-\frac{2M}{\rho}\right)\,\left(\frac{2M}{\rho}\right)f^{(6)}_0&=& \frac{2}{3}\left(\frac{M}{\rho}\right)^2\left(1-\frac{2M}{\rho}\right)f^{(4)}_{0\; ,tt} \nonumber \\
&-& \frac{2}{15}\left(\frac{M}{\rho}\right)^2\left(1-\frac{2M}{\rho}\right)f^{(4)}_{2\; ,tt} \nonumber \\
&-&\left(\frac{M}{\rho}\right)^2\,\Delta^2 f^{(4)}_0\nonumber \\
&-&D_4\,f^{(2)}_{0\; ,tt}
\end{eqnarray}

\begin{eqnarray}\label{f62n}
\,\Box f^{(6)}_{2}+\frac{1}{\rho^2}\,\left(1-\frac{2M}{\rho}\right)\,\left(6+\frac{2M}{\rho}\right)f^{(6)}_2&=& -\frac{2}{3}\left(\frac{M}{\rho}\right)^2\left(1-\frac{2M}{\rho}\right)f^{(4)}_{0\; ,tt} \nonumber \\
&+& \frac{10}{21}\left(\frac{M}{\rho}\right)^2\left(1-\frac{2M}{\rho}\right)f^{(4)}_{2\; ,tt} \nonumber \\
&-& \frac{4}{21}\left(\frac{M}{\rho}\right)^2\left(1-\frac{2M}{\rho}\right)f^{(4)}_{4\; ,tt} \nonumber \\
&-&\left(\frac{M}{\rho}\right)^2\,\Delta^2 f^{(4)}_2\nonumber \\
&-&D_4\,f^{(2)}_{2\; ,tt}-D_6\,f^{(0)}_{2\; ,tt}
\end{eqnarray}

\begin{eqnarray}\label{f64n}
\,\Box f^{(6)}_{4}+\frac{1}{\rho^2}\,\left(1-\frac{2M}{\rho}\right)\,\left(20+\frac{2M}{\rho}\right)f^{(6)}_4&=& - \frac{12}{35}\left(\frac{M}{\rho}\right)^2\left(1-\frac{2M}{\rho}\right)f^{(4)}_{2\; ,tt} \nonumber \\
&+& \frac{38}{77}\left(\frac{M}{\rho}\right)^2\left(1-\frac{2M}{\rho}\right)f^{(4)}_{4\; ,tt} \nonumber \\
&-& \frac{30}{143}\left(\frac{M}{\rho}\right)^2\left(1-\frac{2M}{\rho}\right)f^{(4)}_{6\; ,tt} \nonumber \\
&-&\left(\frac{M}{\rho}\right)^2\,\Delta^2 f^{(4)}_4\nonumber \\
&-&D_4\,f^{(2)}_{4\; ,tt}
\end{eqnarray}

\begin{eqnarray}\label{f66n}
\,\Box f^{(6)}_{6}+\frac{1}{\rho^2}\,\left(1-\frac{2M}{\rho}\right)\,\left(42+\frac{2M}{\rho}\right)f^{(6)}_6&=& - \frac{10}{33}\left(\frac{M}{\rho}\right)^2\left(1-\frac{2M}{\rho}\right)f^{(4)}_{4\; ,tt} \nonumber \\
&+& \frac{82}{165}\left(\frac{M}{\rho}\right)^2\left(1-\frac{2M}{\rho}\right)f^{(4)}_{6\; ,tt} \nonumber \\
&-&\left(\frac{M}{\rho}\right)^2\,\Delta^2 f^{(4)}_6
\end{eqnarray}

\begin{eqnarray}\label{f68n}
\,\Box f^{(6)}_{8}+\frac{1}{\rho^2}\,\left(1-\frac{2M}{\rho}\right)\,\left(72+\frac{2M}{\rho}\right)f^{(6)}_{8}= - \frac{56}{195}\left(\frac{M}{\rho}\right)^2\left(1-\frac{2M}{\rho}\right)f^{(4)}_{6\; ,tt} 
\end{eqnarray}

\begin{eqnarray}\label{f80n}
\,\Box f^{(8)}_{0}+\frac{1}{\rho^2}\,\left(1-\frac{2M}{\rho}\right)\,\left(\frac{2M}{\rho}\right)f^{(8)}_0&=& \frac{2}{3}\left(\frac{M}{\rho}\right)^2\left(1-\frac{2M}{\rho}\right)f^{(6)}_{0\; ,tt} \nonumber \\
&-& \frac{2}{15}\left(\frac{M}{\rho}\right)^2\left(1-\frac{2M}{\rho}\right)f^{(6)}_{2\; ,tt} \nonumber \\
&-&\left(\frac{M}{\rho}\right)^2\,\Delta^2 f^{(6)}_0\nonumber \\
&-&D_4\,f^{(4)}_{0\; ,tt} - D_6\,f^{(2)}_{0\; ,tt} 
\end{eqnarray}

\begin{eqnarray}\label{f82n}
\,\Box f^{(8)}_{2}+\frac{1}{\rho^2}\,\left(1-\frac{2M}{\rho}\right)\,\left(6+\frac{2M}{\rho}\right)f^{(8)}_2&=& -\frac{2}{3}\left(\frac{M}{\rho}\right)^2\left(1-\frac{2M}{\rho}\right)f^{(6)}_{0\; ,tt} \nonumber \\
&+& \frac{10}{21}\left(\frac{M}{\rho}\right)^2\left(1-\frac{2M}{\rho}\right)f^{(6)}_{2\; ,tt} \nonumber \\
&-& \frac{4}{21}\left(\frac{M}{\rho}\right)^2\left(1-\frac{2M}{\rho}\right)f^{(6)}_{4\; ,tt} \nonumber \\
&-&\left(\frac{M}{\rho}\right)^2\,\Delta^2 f^{(6)}_2\nonumber \\
&-&D_4\,f^{(4)}_{2\; ,tt}-D_6\,f^{(2)}_{2\; ,tt} \nonumber \\
&-&D_8\,f^{(0)}_{2\; ,tt}
\end{eqnarray}

\begin{eqnarray}\label{f84n}
\,\Box f^{(8)}_{4}+\frac{1}{\rho^2}\,\left(1-\frac{2M}{\rho}\right)\,\left(20+\frac{2M}{\rho}\right)f^{(8)}_4&=& - \frac{12}{35}\left(\frac{M}{\rho}\right)^2\left(1-\frac{2M}{\rho}\right)f^{(6)}_{2\; ,tt} \nonumber \\
&+& \frac{38}{77}\left(\frac{M}{\rho}\right)^2\left(1-\frac{2M}{\rho}\right)f^{(6)}_{4\; ,tt} \nonumber \\
&-& \frac{30}{143}\left(\frac{M}{\rho}\right)^2\left(1-\frac{2M}{\rho}\right)f^{(6)}_{6\; ,tt} \nonumber \\
&-&\left(\frac{M}{\rho}\right)^2\,\Delta^2 f^{(6)}_4\nonumber \\
&-&D_4\,f^{(4)}_{4\; ,tt}  -D_6\,f^{(2)}_{4\; ,tt}
\end{eqnarray}

\begin{eqnarray}\label{f86n}
\,\Box f^{(8)}_{6}+\frac{1}{\rho^2}\,\left(1-\frac{2M}{\rho}\right)\,\left(42+\frac{2M}{\rho}\right)f^{(8)}_6&=& - \frac{10}{33}\left(\frac{M}{\rho}\right)^2\left(1-\frac{2M}{\rho}\right)f^{(6)}_{4\; ,tt} \nonumber \\ 
&+& \frac{82}{165}\left(\frac{M}{\rho}\right)^2\left(1-\frac{2M}{\rho}\right)f^{(6)}_{6\; ,tt} \nonumber \\
&-&  \frac{56}{255}\left(\frac{M}{\rho}\right)^2\left(1-\frac{2M}{\rho}\right)f^{(6)}_{8\; ,tt} \nonumber \\ 
&-&\left(\frac{M}{\rho}\right)^2\,\Delta^2 f^{(6)}_6-D_4\,f^{(4)}_{6\; ,tt}
\end{eqnarray}

\begin{eqnarray}\label{f88n}
\,\Box f^{(8)}_{8}+\frac{1}{\rho^2}\,\left(1-\frac{2M}{\rho}\right)\,\left(72+\frac{2M}{\rho}\right)f^{(8)}_8&=& - \frac{56}{195}\left(\frac{M}{\rho}\right)^2\left(1-\frac{2M}{\rho}\right)f^{(6)}_{6\; ,tt} \nonumber \\
&+& \frac{142}{285}\left(\frac{M}{\rho}\right)^2\left(1-\frac{2M}{\rho}\right)f^{(6)}_{8\; ,tt} \nonumber \\
&-&\left(\frac{M}{\rho}\right)^2\,\Delta^2 f^{(6)}_8
\end{eqnarray}

\begin{eqnarray}\label{f8_10n}
\,\Box f^{(8)}_{10}+\frac{1}{\rho^2}\,\left(1-\frac{2M}{\rho}\right)\,\left(110+\frac{2M}{\rho}\right)f^{(8)}_{10}= - \frac{90}{323}\left(\frac{M}{\rho}\right)^2\left(1-\frac{2M}{\rho}\right)f^{(6)}_{8\; ,tt} 
\end{eqnarray}

\section{Equations for the partial fields $f^{(6)}_{2\;[k]}$}\label{app_partial_fields}

The partial differential equations for the partial fields $f^{(6)}_{2\;[k]}$ are:

\begin{eqnarray}\label{f62n0}
\,\Box f^{(6)}_{2\;[0]}+\frac{1}{\rho^2}\,\left(1-\frac{2M}{\rho}\right)\,\left(6+\frac{2M}{\rho}\right)f^{(6)}_{2\;[0]}= 0
%-\frac{2}{3}\left(\frac{M}{\rho}\right)^2\left(1-\frac{2M}{\rho}\right)f^{(4)}_{0\; ,tt} %\nonumber \\
%&+& \frac{10}{21}\left(\frac{M}{\rho}\right)^2\left(1-\frac{2M}{\rho}\right)f^{(4)}_{2\; ,tt} \nonumber \\
%&-& \frac{4}{21}\left(\frac{M}{\rho}\right)^2\left(1-\frac{2M}{\rho}\right)f^{(4)}_{4\; ,tt} \nonumber \\
%&-&\left(\frac{M}{\rho}\right)^2\,\Delta^2 f^{(4)}_2\nonumber \\
%&-&D_4\,f^{(2)}_{2\; ,tt}-D_6\,f^{(0)}_{2\; ,tt}
\end{eqnarray}

\begin{eqnarray}\label{f62n1}
\,\Box f^{(6)}_{2\;[1]}+\frac{1}{\rho^2}\,\left(1-\frac{2M}{\rho}\right)\,\left(6+\frac{2M}{\rho}\right)f^{(6)}_{2\;[1]}= -\frac{2}{3}\left(\frac{M}{\rho}\right)^2\left(1-\frac{2M}{\rho}\right)f^{(4)}_{0\; ,tt} %\nonumber \\
%&+& \frac{10}{21}\left(\frac{M}{\rho}\right)^2\left(1-\frac{2M}{\rho}\right)f^{(4)}_{2\; ,tt} \nonumber \\
%&-& \frac{4}{21}\left(\frac{M}{\rho}\right)^2\left(1-\frac{2M}{\rho}\right)f^{(4)}_{4\; ,tt} \nonumber \\
%&-&\left(\frac{M}{\rho}\right)^2\,\Delta^2 f^{(4)}_2\nonumber \\
%&-&D_4\,f^{(2)}_{2\; ,tt}-D_6\,f^{(0)}_{2\; ,tt}
\end{eqnarray}

\begin{eqnarray}\label{f62n2}
\,\Box f^{(6)}_{2\;[2]}+\frac{1}{\rho^2}\,\left(1-\frac{2M}{\rho}\right)\,\left(6+\frac{2M}{\rho}\right)f^{(6)}_{2\;[2]}%&=& -\frac{2}{3}\left(\frac{M}{\rho}\right)^2\left(1-\frac{2M}{\rho}\right)f^{(4)}_{0\; ,tt} \nonumber \\
%&+& 
=\frac{10}{21}\left(\frac{M}{\rho}\right)^2\left(1-\frac{2M}{\rho}\right)f^{(4)}_{2\; ,tt} %\nonumber \\
%&-& \frac{4}{21}\left(\frac{M}{\rho}\right)^2\left(1-\frac{2M}{\rho}\right)f^{(4)}_{4\; ,tt} \nonumber \\
%&-&\left(\frac{M}{\rho}\right)^2\,\Delta^2 f^{(4)}_2\nonumber \\
%&-&D_4\,f^{(2)}_{2\; ,tt}-D_6\,f^{(0)}_{2\; ,tt}
\end{eqnarray}

\begin{eqnarray}\label{f62n3}
\,\Box f^{(6)}_{2\;[3]}+\frac{1}{\rho^2}\,\left(1-\frac{2M}{\rho}\right)\,\left(6+\frac{2M}{\rho}\right)f^{(6)}_{2\;[3]}%&=& -\frac{2}{3}\left(\frac{M}{\rho}\right)^2\left(1-\frac{2M}{\rho}\right)f^{(4)}_{0\; ,tt} \nonumber \\
%&+& 
%=\frac{10}{21}\left(\frac{M}{\rho}\right)^2\left(1-\frac{2M}{\rho}\right)f^{(4)}_{2\; ,tt} %\nonumber \\
%&-& 
=-\frac{4}{21}\left(\frac{M}{\rho}\right)^2\left(1-\frac{2M}{\rho}\right)f^{(4)}_{4\; ,tt} %\nonumber \\
%&-&\left(\frac{M}{\rho}\right)^2\,\Delta^2 f^{(4)}_2\nonumber \\
%&-&D_4\,f^{(2)}_{2\; ,tt}-D_6\,f^{(0)}_{2\; ,tt}
\end{eqnarray}

\begin{eqnarray}\label{f62n4}
\,\Box f^{(6)}_{2\;[4]}+\frac{1}{\rho^2}\,\left(1-\frac{2M}{\rho}\right)\,\left(6+\frac{2M}{\rho}\right)f^{(6)}_{2\;[4]}%&=& -\frac{2}{3}\left(\frac{M}{\rho}\right)^2\left(1-\frac{2M}{\rho}\right)f^{(4)}_{0\; ,tt} \nonumber \\
%&+& 
%=\frac{10}{21}\left(\frac{M}{\rho}\right)^2\left(1-\frac{2M}{\rho}\right)f^{(4)}_{2\; ,tt} %\nonumber \\
%&-& -\frac{4}{21}\left(\frac{M}{\rho}\right)^2\left(1-\frac{2M}{\rho}\right)f^{(4)}_{4\; ,tt} %\nonumber \\
%&-&
=-\left(\frac{M}{\rho}\right)^2\,\Delta^2 f^{(4)}_2%\nonumber \\
%&-&D_4\,f^{(2)}_{2\; ,tt}-D_6\,f^{(0)}_{2\; ,tt}
\end{eqnarray}

\begin{eqnarray}\label{f62n5}
\,\Box f^{(6)}_{2\;[5]}+\frac{1}{\rho^2}\,\left(1-\frac{2M}{\rho}\right)\,\left(6+\frac{2M}{\rho}\right)f^{(6)}_{2\;[5]}%&=& -\frac{2}{3}\left(\frac{M}{\rho}\right)^2\left(1-\frac{2M}{\rho}\right)f^{(4)}_{0\; ,tt} \nonumber \\
%&+& 
%=\frac{10}{21}\left(\frac{M}{\rho}\right)^2\left(1-\frac{2M}{\rho}\right)f^{(4)}_{2\; ,tt} %\nonumber \\
%&-& -\frac{4}{21}\left(\frac{M}{\rho}\right)^2\left(1-\frac{2M}{\rho}\right)f^{(4)}_{4\; ,tt} %\nonumber \\
%&-&
%-\left(\frac{M}{\rho}\right)^2\,\Delta^2 f^{(4)}_2%\nonumber \\
%&-&
=-D_4\,f^{(2)}_{2\; ,tt}%-D_6\,f^{(0)}_{2\; ,tt}
\end{eqnarray}

\begin{eqnarray}\label{f62n6}
\,\Box f^{(6)}_{2\;[6]}+\frac{1}{\rho^2}\,\left(1-\frac{2M}{\rho}\right)\,\left(6+\frac{2M}{\rho}\right)f^{(6)}_{2\;[6]}%&=& -\frac{2}{3}\left(\frac{M}{\rho}\right)^2\left(1-\frac{2M}{\rho}\right)f^{(4)}_{0\; ,tt} \nonumber \\
%&+& 
%=\frac{10}{21}\left(\frac{M}{\rho}\right)^2\left(1-\frac{2M}{\rho}\right)f^{(4)}_{2\; ,tt} %\nonumber \\
%&-& -\frac{4}{21}\left(\frac{M}{\rho}\right)^2\left(1-\frac{2M}{\rho}\right)f^{(4)}_{4\; ,tt} %\nonumber \\
%&-&
%-\left(\frac{M}{\rho}\right)^2\,\Delta^2 f^{(4)}_2%\nonumber \\
%&-&
%-D_4\,f^{(2)}_{2\; ,tt}%
=-D_6\,f^{(0)}_{2\; ,tt}\, .
\end{eqnarray}

For each $f^{(6)}_{2\;[k]}$ partial field the initial datas are zero field and zero time derivative. It is only the non-zero source term that makes the fields not be identically zero. 

We further separate the four different terms in $\,\Delta^2$ in Eq.~(\ref{f62n4}) as follows (see Eq.~(\ref{Delta2_def})), 
\begin{equation}
f^{(6)}_{2\;[4]}:=
\sum_{j=1}^{4}f^{(6)}_{2\;[4(j)]}\, ,
\end{equation}
so that explicitly
\begin{eqnarray}\label{f62n4_1}
\,\Box f^{(6)}_{2\;[4(1)]}+\frac{1}{\rho^2}\,\left(1-\frac{2M}{\rho}\right)\,\left(6+\frac{2M}{\rho}\right)f^{(6)}_{2\;[4]}%&=& -\frac{2}{3}\left(\frac{M}{\rho}\right)^2\left(1-\frac{2M}{\rho}\right)f^{(4)}_{0\; ,tt} \nonumber \\
%&+& 
%=\frac{10}{21}\left(\frac{M}{\rho}\right)^2\left(1-\frac{2M}{\rho}\right)f^{(4)}_{2\; ,tt} %\nonumber \\
%&-& -\frac{4}{21}\left(\frac{M}{\rho}\right)^2\left(1-\frac{2M}{\rho}\right)f^{(4)}_{4\; ,tt} %\nonumber \\
%&-&
=-\left(\frac{M}{\rho}\right)^2\,\Delta_{(1)}^2 f^{(4)}_2%\nonumber \\
%&-&D_4\,f^{(2)}_{2\; ,tt}-D_6\,f^{(0)}_{2\; ,tt}
\end{eqnarray}
\begin{eqnarray}\label{f62n4_2}
\,\Box f^{(6)}_{2\;[4(2)]}+\frac{1}{\rho^2}\,\left(1-\frac{2M}{\rho}\right)\,\left(6+\frac{2M}{\rho}\right)f^{(6)}_{2\;[4]}%&=& -\frac{2}{3}\left(\frac{M}{\rho}\right)^2\left(1-\frac{2M}{\rho}\right)f^{(4)}_{0\; ,tt} \nonumber \\
%&+& 
%=\frac{10}{21}\left(\frac{M}{\rho}\right)^2\left(1-\frac{2M}{\rho}\right)f^{(4)}_{2\; ,tt} %\nonumber \\
%&-& -\frac{4}{21}\left(\frac{M}{\rho}\right)^2\left(1-\frac{2M}{\rho}\right)f^{(4)}_{4\; ,tt} %\nonumber \\
%&-&
=-\left(\frac{M}{\rho}\right)^2\,\Delta_{(2)}^2 f^{(4)}_2%\nonumber \\
%&-&D_4\,f^{(2)}_{2\; ,tt}-D_6\,f^{(0)}_{2\; ,tt}
\end{eqnarray}
\begin{eqnarray}\label{f62n4_3}
\,\Box f^{(6)}_{2\;[4(3)]}+\frac{1}{\rho^2}\,\left(1-\frac{2M}{\rho}\right)\,\left(6+\frac{2M}{\rho}\right)f^{(6)}_{2\;[4]}%&=& -\frac{2}{3}\left(\frac{M}{\rho}\right)^2\left(1-\frac{2M}{\rho}\right)f^{(4)}_{0\; ,tt} \nonumber \\
%&+& 
%=\frac{10}{21}\left(\frac{M}{\rho}\right)^2\left(1-\frac{2M}{\rho}\right)f^{(4)}_{2\; ,tt} %\nonumber \\
%&-& -\frac{4}{21}\left(\frac{M}{\rho}\right)^2\left(1-\frac{2M}{\rho}\right)f^{(4)}_{4\; ,tt} %\nonumber \\
%&-&
=-\left(\frac{M}{\rho}\right)^2\,\Delta_{(3)}^2 f^{(4)}_2%\nonumber \\
%&-&D_4\,f^{(2)}_{2\; ,tt}-D_6\,f^{(0)}_{2\; ,tt}
\end{eqnarray}
\begin{eqnarray}\label{f62n4_4}
\,\Box f^{(6)}_{2\;[4(4)]}+\frac{1}{\rho^2}\,\left(1-\frac{2M}{\rho}\right)\,\left(6+\frac{2M}{\rho}\right)f^{(6)}_{2\;[4]}%&=& -\frac{2}{3}\left(\frac{M}{\rho}\right)^2\left(1-\frac{2M}{\rho}\right)f^{(4)}_{0\; ,tt} \nonumber \\
%&+& 
%=\frac{10}{21}\left(\frac{M}{\rho}\right)^2\left(1-\frac{2M}{\rho}\right)f^{(4)}_{2\; ,tt} %\nonumber \\
%&-& -\frac{4}{21}\left(\frac{M}{\rho}\right)^2\left(1-\frac{2M}{\rho}\right)f^{(4)}_{4\; ,tt} %\nonumber \\
%&-&
=-\left(\frac{M}{\rho}\right)^2\,\Delta_{(4)}^2 f^{(4)}_2%\nonumber \\
%&-&D_4\,f^{(2)}_{2\; ,tt}-D_6\,f^{(0)}_{2\; ,tt}
\end{eqnarray}
where
\begin{eqnarray}
\Delta_{(1)}^2 &:=& - \frac{2M}{\rho}\frac{1}{1-\frac{M}{\rho}}\,\partial^2_t
%+\frac{1}{\left(1-\frac{M}{\rho}\right)^2}\,\partial^2_{\rho_*}
%-\frac{1}{\rho}\frac{1-\frac{2M}{\rho}}{\left(1-\frac{M}{\rho}\right)^3}\,\partial_{\rho_*}
%\nonumber\\
%&+&\frac{1}{\rho^2}\left(1-\frac{4M}{\rho}+\frac{2M^2}{\rho^2}\right)\,\frac{1-\frac{2M}{\rho}}{\left(1-\frac{M}{\rho}\right)^3} \, ,
\end{eqnarray}
\begin{eqnarray}
\Delta_{(2)}^2 &:=&% - \frac{2M}{\rho}\frac{1}{1-\frac{M}{\rho}}\,\partial^2_t +
\frac{1}{\left(1-\frac{M}{\rho}\right)^2}\,\partial^2_{\rho_*}
%-\frac{1}{\rho}\frac{1-\frac{2M}{\rho}}{\left(1-\frac{M}{\rho}\right)^3}\,\partial_{\rho_*}
%\nonumber\\
%&+&\frac{1}{\rho^2}\left(1-\frac{4M}{\rho}+\frac{2M^2}{\rho^2}\right)\,\frac{1-\frac{2M}{\rho}}{\left(1-\frac{M}{\rho}\right)^3} \, ,
\end{eqnarray}
\begin{eqnarray}
\Delta_{(3)}^2 &:=&% - \frac{2M}{\rho}\frac{1}{1-\frac{M}{\rho}}\,\partial^2_t +
%\frac{1}{\left(1-\frac{M}{\rho}\right)^2}\,\partial^2_{\rho_*}
-\frac{1}{\rho}\frac{1-\frac{2M}{\rho}}{\left(1-\frac{M}{\rho}\right)^3}\,\partial_{\rho_*}
%\nonumber\\
%&+&\frac{1}{\rho^2}\left(1-\frac{4M}{\rho}+\frac{2M^2}{\rho^2}\right)\,\frac{1-\frac{2M}{\rho}}{\left(1-\frac{M}{\rho}\right)^3} \, ,
\end{eqnarray}
\begin{eqnarray}
\Delta_{(4)}^2 &:=&% - \frac{2M}{\rho}\frac{1}{1-\frac{M}{\rho}}\,\partial^2_t +
%\frac{1}{\left(1-\frac{M}{\rho}\right)^2}\,\partial^2_{\rho_*}
%-\frac{1}{\rho}\frac{1-\frac{2M}{\rho}}{\left(1-\frac{M}{\rho}\right)^3}\,\partial_{\rho_*}
%\nonumber\\
\frac{1}{\rho^2}\left(1-\frac{4M}{\rho}+\frac{2M^2}{\rho^2}\right)\,\frac{1-\frac{2M}{\rho}}{\left(1-\frac{M}{\rho}\right)^3} \, .
\end{eqnarray}

\section{The dominant channel for $\ell'=4$, $\ell=4$}\label{app:44}
For the case $\ell'=4$, $\ell=4$ the field equations are:
\begin{eqnarray}\label{f04n}
\,\Box f^{(0)}_{4}+\frac{1}{\rho^2}\,\left(1-\frac{2M}{\rho}\right)\,\left(20+\frac{2M}{\rho}\right)f^{(0)}_4=0 \end{eqnarray}
\begin{eqnarray}\label{f22n}
\,\Box f^{(2)}_{2}+\frac{1}{\rho^2}\,\left(1-\frac{2M}{\rho}\right)\,\left(6+\frac{2M}{\rho}\right)f^{(2)}_2=-\frac{4}{21}\left(\frac{M}{\rho}\right)^2\left(1-\frac{2M}{\rho}\right)f^{(0)}_{4\; ,tt}\end{eqnarray}
\begin{eqnarray}\label{f40n}
\,\Box f^{(4)}_{0}+\frac{1}{\rho^2}\,\left(1-\frac{2M}{\rho}\right)\,\left(\frac{2M}{\rho}\right)f^{(4)}_0=-\frac{2}{15}\left(\frac{M}{\rho}\right)^2\left(1-\frac{2M}{\rho}\right)f^{(2)}_{2\; ,tt} \end{eqnarray}
\begin{eqnarray}\label{f6n}
\,\Box f^{(6)}_{2}+\frac{1}{\rho^2}\,\left(1-\frac{2M}{\rho}\right)\,\left(6+\frac{2M}{\rho}\right)f^{(6)}_2=- \frac{2}{3}\left(\frac{M}{\rho}\right)^2\left(1-\frac{2M}{\rho}\right)f^{(4)}_{0\; ,tt} 
\nonumber \\+\frac{1}{2}\left(\frac{M}{\rho}\right)^5\,\frac{1-\frac{2M}{\rho}+\frac{2M^2}{\rho^2}}{\left(1-\frac{M}{\rho}\right)^3}f^{(2)}_{2\; ,tt} 
\end{eqnarray}
\begin{eqnarray}\label{f8n}
\,\Box f^{(8)}_{4}+\frac{1}{\rho^2}\,\left(1-\frac{2M}{\rho}\right)\,\left(20+\frac{2M}{\rho}\right)f^{(8)}_4=- \frac{12}{35}\left(\frac{M}{\rho}\right)^2\left(1-\frac{2M}{\rho}\right)f^{(6)}_{2\; ,tt} \nonumber
\\+\frac{5}{32}\left(\frac{M}{\rho}\right)^{9}\,\frac{1-\frac{2M}{\rho}+\frac{2M^2}{\rho^2}}{\left(1-\frac{M}{\rho}\right)^7}f^{(0)}_{4\; ,tt} \, .
\end{eqnarray}

\section{The dominant channel for $\ell'=6$, $\ell=0,2,4,6$}\label{app:66}

The evolution equations are
\begin{eqnarray}\label{f06}
\,\Box f^{(0)}_{6}+\frac{1}{\rho^2}\,\left(1-\frac{2M}{\rho}\right)\,\left(42+\frac{2M}{\rho}\right)f^{(0)}_6=0 \end{eqnarray}
\begin{eqnarray}\label{f24}
\,\Box f^{(2)}_{4}+\frac{1}{\rho^2}\,\left(1-\frac{2M}{\rho}\right)\,\left(20+\frac{2M}{\rho}\right)f^{(2)}_4=-\frac{30}{143}\left(\frac{M}{\rho}\right)^2\left(1-\frac{2M}{\rho}\right)f^{(0)}_{6\; ,tt}\end{eqnarray}
\begin{eqnarray}\label{f42}
\,\Box f^{(4)}_{2}+\frac{1}{\rho^2}\,\left(1-\frac{2M}{\rho}\right)\,\left(6+\frac{2M}{\rho}\right)f^{(4)}_2=-\frac{4}{21}\left(\frac{M}{\rho}\right)^2\left(1-\frac{2M}{\rho}\right)f^{(2)}_{4\; ,tt}\end{eqnarray}
\begin{eqnarray}\label{f60}
\,\Box f^{(6)}_{0}+\frac{1}{\rho^2}\,\left(1-\frac{2M}{\rho}\right)\,\left(\frac{2M}{\rho}\right)f^{(6)}_0=-\frac{2}{15}\left(\frac{M}{\rho}\right)^2\left(1-\frac{2M}{\rho}\right)f^{(4)}_{2\; ,tt} \end{eqnarray}
\begin{eqnarray}\label{f82}
\,\Box f^{(8)}_{2}+\frac{1}{\rho^2}\,\left(1-\frac{2M}{\rho}\right)\,\left(6+\frac{2M}{\rho}\right)f^{(8)}_2=- \frac{2}{3}\left(\frac{M}{\rho}\right)^2\left(1-\frac{2M}{\rho}\right)f^{(6)}_{0\; ,tt} 
\nonumber \\+\frac{1}{2}\left(\frac{M}{\rho}\right)^5\,\frac{1-\frac{2M}{\rho}+\frac{2M^2}{\rho^2}}{\left(1-\frac{M}{\rho}\right)^3}f^{(4)}_{2\; ,tt} 
\end{eqnarray}
\begin{eqnarray}\label{f8n}
\,\Box f^{(10)}_{4}+\frac{1}{\rho^2}\,\left(1-\frac{2M}{\rho}\right)\,\left(20+\frac{2M}{\rho}\right)f^{(10)}_4=- \frac{12}{35}\left(\frac{M}{\rho}\right)^2\left(1-\frac{2M}{\rho}\right)f^{(8)}_{2\; ,tt} \nonumber
\\+\frac{5}{32}\left(\frac{M}{\rho}\right)^{9}\,\frac{1-\frac{2M}{\rho}+\frac{2M^2}{\rho^2}}{\left(1-\frac{M}{\rho}\right)^7}f^{(2)}_{4\; ,tt} 
\end{eqnarray}
\begin{eqnarray}\label{f8n}
\,\Box f^{(12)}_{6}+\frac{1}{\rho^2}\,\left(1-\frac{2M}{\rho}\right)\,\left(42+\frac{2M}{\rho}\right)f^{(12)}_6=- \frac{10}{33}\left(\frac{M}{\rho}\right)^2\left(1-\frac{2M}{\rho}\right)f^{(10)}_{4\; ,tt} \nonumber
\\+\frac{21}{256}\left(\frac{M}{\rho}\right)^{13}\,\frac{1-\frac{2M}{\rho}+\frac{2M^2}{\rho^2}}{\left(1-\frac{M}{\rho}\right)^{11}}f^{(0)}_{6\; ,tt} 
\end{eqnarray}

\section{The dominant channel for $\ell'=2$, $\ell=0,2,4,6$}\label{app:26}

Next, we consider $\ell'=2$, $\ell=4$ or $\ell=6$. For the channel of interest we take

and the evolution equations are
\begin{eqnarray}\label{f02}
\,\Box f^{(0)}_{2}+\frac{1}{\rho^2}\,\left(1-\frac{2M}{\rho}\right)\,\left(6+\frac{2M}{\rho}\right)f^{(0)}_2=0 \end{eqnarray}
\begin{eqnarray}\label{f20}
\,\Box f^{(2)}_{0}+\frac{1}{\rho^2}\,\left(1-\frac{2M}{\rho}\right)\,\left(\frac{2M}{\rho}\right)f^{(2)}_0=-\frac{2}{15}\left(\frac{M}{\rho}\right)^2\left(1-\frac{2M}{\rho}\right)f^{(0)}_{2\; ,tt} \end{eqnarray}
\begin{eqnarray}\label{f42}
\,\Box f^{(4)}_{2}+\frac{1}{\rho^2}\,\left(1-\frac{2M}{\rho}\right)\,\left(6+\frac{2M}{\rho}\right)f^{(4)}_2=- \frac{2}{3}\left(\frac{M}{\rho}\right)^2\left(1-\frac{2M}{\rho}\right)f^{(2)}_{0\; ,tt} 
\nonumber \\+\frac{1}{2}\left(\frac{M}{\rho}\right)^5\,\frac{1-\frac{2M}{\rho}+\frac{2M^2}{\rho^2}}{\left(1-\frac{M}{\rho}\right)^3}f^{(0)}_{2\; ,tt} 
\end{eqnarray}
\begin{eqnarray}\label{f64}
\,\Box f^{(6)}_{4}+\frac{1}{\rho^2}\,\left(1-\frac{2M}{\rho}\right)\,\left(20+\frac{2M}{\rho}\right)f^{(6)}_4=- \frac{12}{35}\left(\frac{M}{\rho}\right)^2\left(1-\frac{2M}{\rho}\right)f^{(4)}_{2\; ,tt} 
\end{eqnarray}
\begin{eqnarray}\label{f86}
\,\Box f^{(8)}_{6}+\frac{1}{\rho^2}\,\left(1-\frac{2M}{\rho}\right)\,\left(42+\frac{2M}{\rho}\right)f^{(8)}_6=- \frac{10}{33}\left(\frac{M}{\rho}\right)^2\left(1-\frac{2M}{\rho}\right)f^{(6)}_{4\; ,tt} 
\end{eqnarray}

\end{appendix}

\end{document}